\documentclass[twocolumn,aps,prl,showpacs,amsmath,amssymb,superscriptaddress, longbibliography]{revtex4-1}

\usepackage{physics}
\usepackage{graphicx} 
\usepackage[caption=false, labelformat=simple]{subfig}
\usepackage{dcolumn}
\usepackage{csquotes}
\usepackage[colorlinks=true,linkcolor = {blue}, citecolor = {blue}, urlcolor = {blue}]{hyperref}
\usepackage{xcolor}

\usepackage[export]{adjustbox}

\begin{document}

\title{Probing non-Gaussian correlations through entanglement generation in a many-body quantum system}

\author{J. van de Kraats}
\email[Corresponding author:  ]{j.v.d.kraats@tue.nl}
\affiliation{Department of Applied Physics and Science Education, Eindhoven University of Technology, P. O. Box 513, 5600 MB Eindhoven, The Netherlands}
\author{D.J.M. Ahmed-Braun}
\affiliation{TQC, Departement Fysica, Universiteit Antwerpen, Universiteitsplein 1, B-2610 Antwerpen, Belgium}
\author{V.E. Colussi}
\affiliation{Quantinuum, 303 S Technology Ct, Broomfield, CO 80021, USA}
\author{S.J.J.M.F. Kokkelmans}
\affiliation{Department of Applied Physics and Science Education, Eindhoven University of Technology, P. O. Box 513, 5600 MB Eindhoven, The Netherlands}
\date{\today}

\begin{abstract}
In understanding strongly correlated quantum systems, quantifying the non-Gaussian nature of interparticle correlations is invaluable. We show that, for a uniform quantum gas, there exists a natural connection between non-Gaussian correlations and the generation of momentum-space entanglement. Furthermore, this entanglement can be directly measured in an experiment using time-of-flight techniques. To prototype our method, we numerically study entanglement generation in a degenerate Bose gas following a quench to the unitary regime, where Gaussian and non-Gaussian correlations are generated sequentially in time.
\end{abstract}

\maketitle

\textit{Introduction.}--- Entanglement has attracted both great interest and considerable controversy since the early days of quantum mechanics, most famously expressed by the ``EPR paradox" which posits that quantum mechanics is inconsistent with local realism \cite{Einstein1935, Reid2009}. Today however, entanglement is well established as an empirical fact, and is regarded as a necessary resource for achieving quantum advantage in next-generation quantum technologies \cite{Acín2018, Preskill2018}. In addition to to these direct applications, the associated development of quantum information science has also sparked interest in using entanglement as a tool for studying many-body systems \cite{Amico2008, Horodecki2009, Hauke2016, Kunkel2022}. For example, entanglement entropy has found use in condensed matter physics as an effective probe for detecting topological order and quantum critical behavior \cite{Osterloh2002, Vidal2003, Kitaev2006, Li2008, Jiang2012}, and understanding thermalization in out-of-equilibrium quantum systems \cite{Eisert2015, Abanin2019, Chang2019, Mueller2022}.

It is well known that generation of entanglement is intimately connected to the presence of non-Gaussian processes \cite{Walschaers2021}. For example, entanglement distillation of Gaussian states is only possible with non-Gaussian operations \cite{Eisert2002}. This fact naturally leads to the question of whether entanglement can be used as a probe for quantifying the presence of non-Gaussian correlations in many-body quantum systems \cite{Strobel2014, Tenart2021, Bureik2025, Froland2024}. Given the significant recent research interest in non-Gaussian processes, for instance in the generation of long-wavelength few-body correlations beyond the quasiparticle picture, such tools will be exceedingly valuable \cite{Armijo2010, Fletcher2017, Klauss2017, Schweigler2017, DeSalvo2019, Mestrom2020, MusolinoPRL, Patel2023, Zhang2023, Bureik2025}. Particularly demanding are continuous variable systems in three dimensions, where current widespread theoretical approaches based on matrix product states and tensor networks become too demanding computationally, especially for studying non-equilibrium dynamics \cite{Haegeman2016, Kloss2018, Dutta2022, Ganahl2023}.
\begin{figure}
\includegraphics{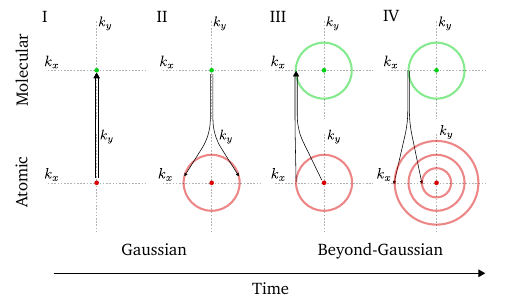}
\caption{\label{fig:GrowthDiag} Schematic representation of the four fundamental steps that generate intershell entanglement in the unitary Bose gas. Each panel shows the two-dimensional plane $k_z = 0$ in the atomic (red) or molecular (green) space of momentum modes. Dots in the center represent the condensates, and shells represent spherically symmetric excitations at nonzero momentum.}
\end{figure}

In this Letter we show that, for a uniform quantum gas, the entanglement entropy between thin shells in momentum space can act as a witness for non-Gaussian processes \cite{Flynn2023}. Furthermore, this entanglement entropy can be directly calculated by measuring density-density correlation functions, which are accessible through commonplace time-of-flight techniques \cite{Altman2004, Greiner2005, Cayla2018, Tenart2021}. We present our formalism using a paradigmatic example of a three-dimensional quantum system containing strong non-Gaussian correlations, namely an ideal degenerate Bose gas quenched to the strongly interacting regime. Since the pioneering experimental works of Refs. \cite{Makotyn2014, Klauss2017, Eigen2017, Eigen2018}, the correlation dynamics in this system has received considerable theoretical interest \cite{Kira2015, Kira2015_2, Regemortel2018, Colussi2018, Colussi2018_2, Munoz2019, Colussi2020, MusolinoPRL, Braun2022, Kraats2024}. Initially, the system is dominated by Gaussian correlations, which induce the formation of a universal prethermal state where macroscopic observables look quasi-equilibrated \cite{Regemortel2018, Colussi2020}. At later times, higher order non-Gaussian correlations develop, predominantly due to three-body scattering processes, which drive the system out of the prethermal stage and towards true thermalization, and generate off-diagonal long-range order in three-body correlations \cite{MusolinoPRL, Kraats2024}. We will show that this temporal separation of Gaussian and non-Gaussian dynamics becomes particularly natural when formulated in terms of momentum space entanglement.

\textit{Model.}--- We consider an ideal homogeneous $N$-particle Bose-Einstein Condensate (BEC) which is instantaneously quenched to the unitary regime, where the two-body s-wave scattering length $a \rightarrow \infty$ and interactions are as strong as allowed by quantum mechanics. The quench is initiated using a magnetic Feshbach resonance, which couples the atoms to a weakly bound diatomic molecular state \cite{Kohler2006, Chin2010}. The subsequent quantum depletion of the condensate generates correlations between the excited momentum modes $\abs*{\vb{k}} > 0$, described by annihilation operators $\hat{a}_{\vb{k}}$. As pointed out in Ref.~\cite{Kraats2024}, the post-quench correlation growth dynamics can be captured by the following two-point and three-point connected correlations functions, or cumulants,
\begin{equation}
n_{\vb{k}} = \expval*{\hat{a}_{\vb{k}}^{\dagger} \hat{a}_{\vb{k}}}, \ \kappa_{\vb{k}} = \expval*{\hat{a}_{\vb{k}} \hat{a}_{-\vb{k}}}, \ R_{\vb{k}, \vb{q}} = \expval*{\hat{a}_{\vb{k}} \hat{a}_{\vb{q}} \hat{a}_{-\vb{k} - \vb{q}}}.
\label{eq:cus}
\end{equation}
Here $n_{\vb{k}}$ is the excitation density, $\kappa_{\vb{k}}$ is the anomalous pairing density, and $R_{\vb{k}, \vb{q}}$ the anomalous tripling density \cite{Colussi2018_2, Colussi2020}. Eventually, as the condensate is depleted and the occupation of excited modes grows, higher order correlations are sequentially generated \cite{Kira2015, Colussi2020}. Hence, the truncation above is justified at early times, where $n_{\vb{k}}$ is small \cite{Kraats2024}. In the widely used Hartree-Fock-Bogoliubov (HFB) approximation \cite{Blazoit1985, Braun2022}, three-body processes in the system are neglected by also setting $R_{\vb{k}, \vb{q}} = 0$, which results in a fully integrable and Gaussian model of the dynamics, appropriate for describing early times following the quench. Upon including the non-Gaussian cumulant $R_{\vb{k}, \vb{q}}$, the additional three-body scattering processes break integrability \cite{Regemortel2018}, and significantly improve the match between model and experiment once the gas departs from the universal prethermal state \cite{Kraats2024}. Additionally, it was recently shown that $R_{\vb{k}, \vb{q}}$ defines an order parameter for a novel Bose-Einstein condensate of correlated triples, which is particularly sensitive to the Efimov effect characteristic for strongly interacting three-body systems \cite{MusolinoPRL}.

\textit{Entanglement in thin shells.}--- We follow the standard formulation of bipartite entanglement by dividing the system into regions $\mathcal{A}$ and $\mathcal{B}$, and quantify the entanglement between the two spaces with the R\'enyi entropy functional \cite{Renyi1961, Horodecki2009},
\begin{equation}
S_{\gamma}(\hat{\rho}) = \frac{1}{1 - \gamma} \ln \mathrm{Tr} \left[ \hat{\rho}^{\gamma} \right] \quad \forall \gamma \in (0, 1) \cup (1, \infty),
\label{eq:Rent}
\end{equation}
where $\hat{\rho} = \mathrm{Tr}_{\mathcal{B}} \left[\dyad{\Psi}\right]$ is the reduced density matrix of the total many-body quantum state $\ket*{\Psi}$. The structure of the relevant correlations in Eq.~\eqref{eq:cus} suggests that a useful bipartition for probing non-Gaussianity is obtained by defining space $\mathcal{A}$ as containing all atomic momentum modes in a thin spherical shell with variable radius $k_s$ and width $\delta k_s$, thus placing all other atomic modes and all molecular modes in space $\mathcal{B}$ \cite{Flynn2023}. This partition naturally separates the Gaussian correlation $\kappa_{\vb{k}}$, which correlates particles \textit{within} a given shell, and the non-Gaussian correlation $R_{\vb{k}, \vb{q}}$, which may correlate particles \textit{between} different shells.

In Fig.~\ref{fig:GrowthDiag} panels I-IV, we illustrate the application of the thin-shell partition to the dynamics of the quenched unitary Bose gas. Initially, association of atoms inside the BEC leads to the formation of a molecular condensate (I) \cite{Romans2004, Radzihovsky2004, Radzihovsky2008, Zhang2021, SupMat}. These molecules may then dissociate into correlated pairs (II), which must end up within a shell that defines a specific thin-shell bipartition. In the Gaussian model, this two-step excitation process is the only pathway for quantum depletion, implying that $\hat{\rho}$ is always a pure state with vanishing R\'enyi entropy. Note here that we assume the $U(1)$ symmetry breaking picture of Bose-Einstein condensation where the atomic and molecular condensates are described by classical fields, and consequently do not entangle with the excited modes \cite{Pitaevskii2016, Frerot2015}. If however, we account for the three-body correlations represented by $R_{\vb{k}, \vb{q}}$, atoms within a shell can interact with the condensate to form excited molecules (III) \cite{Kraats2024}, whose subsequent decay can generate intershell entanglement (IV). This highlights the ergodic nature of this beyond-Gaussian model, and, most importantly for the purposes of this work, indicates how the entanglement entropy in space $\mathcal{A}$ provides a direct measure for the strength of non-Gaussian three-body correlations.

To quantify the entanglement entropy, we note that, within the coherent-state representation, $\hat{\rho}$ can be related directly to the cumulants in Eq.~\eqref{eq:cus} \cite{Blazoit1985, Frerot2015, Kira2008, Serafini2017, SupMat}. In this approach, the shell width $\delta k_s$ acts as a perturbative parameter, as has also been pointed out for interacting Fermi gases in Ref.~\cite{Flynn2023}. Specifically, retaining only the lowest order contributions in $\delta k_s$ gives the Gaussian state $\hat{\rho} = \hat{\rho}_G$, where, 
\begin{equation}
\hat{\rho}_G = \prod_{\vb{k}\in \mathcal{A}} \frac{1}{1 + \tilde{n}_{\vb{k}}} e^{- \lambda_{\vb{k}} \hat{b}_{\vb{k}}^{\dagger} \hat{b}_{\vb{k}}},
\label{eq:rhoG}
\end{equation}
which we will refer to as the \textit{thin-shell approximation}. Here we have introduced the canonical transformation $\hat{a}_{\vb{k}} = u_{\vb{k}} \hat{b}_{\vb{k}} + v_{\vb{k}}^* \hat{b}_{-\vb{k}}^{\dagger}$ in space $\mathcal{A}$, where $u_{\vb{k}}, v_{\vb{k}}$ are known functions of $n_{\vb{k}}$ and $\kappa_{\vb{k}}$ with the symplectic property $\abs*{u_{\vb{k}}}^2 - \abs*{v_{\vb{k}}}^2 = 1$ \cite{SupMat}.. The single-particle states annihilated by $\hat{b}_{\vb{k}}$ are known as entanglement eigenmodes \cite{Frerot2015}, with associated mode occupations $\tilde{n}_{\vb{k}} = \expval*{\hat{b}_{\vb{k}}^{\dagger} \hat{b}_{\vb{k}}} = 1/(e^{\lambda_{\vb{k}}} - 1)$, or, 
\begin{equation}
\tilde{n}_{\vb{k}} = \frac{1}{2} \left(-1 + \sqrt{(1 + 2n_{\vb{k}})^2 - 4 \abs{\kappa_{\vb{k}}}^2} \right).
\label{eq:QPnumber}
\end{equation}
Intuitively, Eq.~\eqref{eq:rhoG} can be interpreted as a thermal state at a temperature $T = 1$ generated by a Hamiltonian $\mathcal{\hat{H}} = \sum_{\vb{k} \in \mathcal{A}} \lambda_{\vb{k}} \hat{b}_{\vb{k}}^{\dagger} \hat{b}_{\vb{k}} $, typically referred to as the entanglement Hamiltonian \cite{Li2008, Dalmonte2022}. Its eigenvalues define the entanglement spectrum of the state, which for a given shell is gapped by the dimensionless ``energy" $\lambda_{\vb{k}}$. Equation~\eqref{eq:QPnumber} indicates that the entanglement spectrum in the thin-shell approximation is fully expressed in terms of the doublet cumulants $n_{\vb{k}}$ and $\abs*{\kappa_{\vb{k}}}$, which greatly simplifies studying the entanglement structure of the state. For example, the R\'enyi entropy follows directly from Eq.~\eqref{eq:Rent} as,
\begin{equation}
S_{\gamma}(\hat{\rho}_G) = \frac{1}{\gamma - 1} \sum_{\vb{k} \in \mathcal{A}} \ln \left[ (1 + \tilde{n}_{\vb{k}})^{\gamma} - \tilde{n}_{\vb{k}}^{\gamma} \right],
\label{eq:Snk}
\end{equation}
which becomes an intensive quantity independent of $\delta k_s$ upon normalizing to the number of momentum modes $M_{\mathcal{A}}$ inside the shell, i.e. $s_{\gamma}(\hat{\rho}_G) \equiv S_{\gamma}(\hat{\rho}_G)/M_{\mathcal{A}}$. 

It is important to understand that, even if non-Gaussian processes exist in the system, the thin-shell state in Eq.~\eqref{eq:rhoG} is always Gaussian in form. The presence of non-Gaussian processes does however affect the spectrum $\lambda_{\vb{k}}$. Equation~\eqref{eq:Snk} suggests that in the Gaussian model, where the entanglement entropy vanishes, it must always hold that $\tilde{n}_{\vb{k}} = 0$, meaning that the entanglement spectrum has an infinite gap. This condition is indeed enforced by the many-body equations of motion \cite{SupMat}, but attains a particularly elegant form in a phase space picture of the quantum state \cite{Kira2015}. Within one half of the momentum shell ($\vb{k}>0$), we introduce the bidirectional basis $\hat{d}_{\vb{k}} = e^{-i \theta_{\vb{k}}} (\hat{a}_{\vb{k}} + \hat{a}_{-\vb{k}})/\sqrt{2}$
and fix the phase $\theta_{\vb{k}}$ such that $\expval*{\hat{d}_{\vb{k}}^{\dagger} \hat{d}_{\vb{k}}} = n_{\vb{k}}$ and $\expval*{\hat{d}_{\vb{k}}\hat{d}_{\vb{k}}} = \abs*{\kappa_{\vb{k}}}$. These operators define Hermitian field quadratures, $\hat{X}_{\vb{k}} =  (\hat{d}_{\vb{k}} + \hat{d}_{\vb{k}}^{\dagger} )/2$, and  $\hat{Y}_{\vb{k}} =  (\hat{d}_{\vb{k}} - \hat{d}_{\vb{k}}^{\dagger} )/2i$, which obey the canonical commutation relation $[\hat{X}_{\vb{k}}, \hat{Y}_{\vb{q}}] = i\delta_{\vb{k}, \vb{q}}/2$. From here we can directly compute the standard deviations $\Delta X_{\vb{k}}$ and $\Delta Y_{\vb{k}}$, with associated Robertson uncertainty relation \cite{Robertson1929},
\begin{equation}
4\Delta X_{\vb{k}} \Delta Y_{\vb{k}} = (1 + 2n_{\vb{k}})^2 - 4 \abs{\kappa_{\vb{k}}}^2 \ge 1.
\label{eq:Unc}
\end{equation}
This inequality can also be derived from the positivity condition for the relative number squeezing parameter for $(\vb{k}, -\vb{k})$ modes, as introduced in Refs.~\cite{Roberts2002, Tenart2021}. It is crucial to note however, that relative number squeezing indicates intrashell entanglement, which may be generated by Gaussian processes. Our thin-shell framework neglects this entanglement by construction, but rather focuses on the intershell entanglement, which arises out of the non-Gaussianity of correlations. Comparing with Eq.~\eqref{eq:QPnumber} we recognize that a vanishing eigenmode occupation, $\tilde{n}_{\vb{k}} = 0$, corresponds with saturation of the uncertainty relation. Hence, $n_{\vb{k}}$ and $\abs*{\kappa_{\vb{k}}}$ together classify a thin-shell state as being: (1) forbidden ($\tilde{n}_{\vb{k}} < 0$), (2) unentangled ($\tilde{n}_{\vb{k}} = 0$) or (3) entangled ($\tilde{n}_{\vb{k}} > 0$).

Before moving on to results, we want to emphasize that the formalism we have developed above presents a clear opportunity for an experiment. By measuring $n_{\vb{k}}$ and $\abs{\kappa_{\vb{k}}}$, both of which are accessible through time-of-flight techniques \cite{Altman2004, Greiner2005, Cayla2018, Tenart2021}, the entanglement spectrum and all derived quantities such as the entanglement entropy follow directly. As we have already argued, and will show in more detail in the remainder of this Letter, the entanglement characterized via this method then provides a direct probe of non-Gaussian processes in the system.

\textit{Results.}--- Having established the formalism for entropy generation in the unitary Bose gas, we will now characterize this entropy using a numerical simulation of the system. We employ the model developed in Ref.~\cite{Kraats2024}, which formulates the many-body dynamics in terms of coupled nonlinear differential equations for a set of relevant atomic and molecular cumulants, including those presented in Eq.~\eqref{eq:cus} (see Supplemental Material \cite{SupMat} for more details). In this model the interaction is characterized by the inverse scattering length $1/a$, which we fix at zero, and the intrinsic length $R_*$ \cite{Petrov2004}, which determines the molecular lifetime and characterizes the Feshbach resonance as broad ($R_* k_n \ll 1$) or narrow ($R_* k_n \gg 1$) \cite{Ho2012}. As we discuss in more detail later, the intrinsic length can also be used to control the validity of the thin-shell approximation. Going forward, we will express all quantities in units of the Fermi momentum $k_n = (6 \pi^2 n)^{1/3}$ and Fermi time $t_n = 2m/k_n^2$, where $n$ is the gas density and $m$ the atomic mass. We limit our simulations to times $t/t_n < 3$, where the model has been shown to give good agreement with experiment \cite{Kraats2024}.

\begin{figure}
\includegraphics{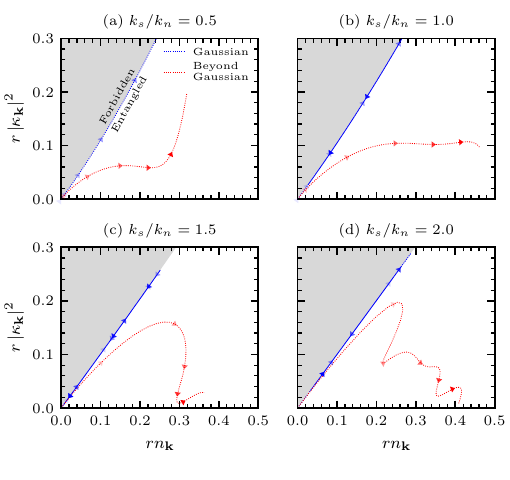}
\caption{\label{fig:Trajectory} Trajectory through $(n_{\vb{k}}, \abs{\kappa_{\vb{k}}}^2)$ phase space for the Gaussian and beyond-Gaussian models in blue and red respectively, for $R_* k_n  =0.4$ and $0 \le t/t_n \le 3$. The gray region is forbidden by Eq.~\eqref{eq:Unc} and the white region is associated with intershell entanglement. Increasingly opaque arrows show the direction of time. For visibility we rescale the axes of the four panels (a)-(d) by magnification factors $r = (1, 2, 10, 50)$.}
\end{figure}

\begin{figure*}
\includegraphics{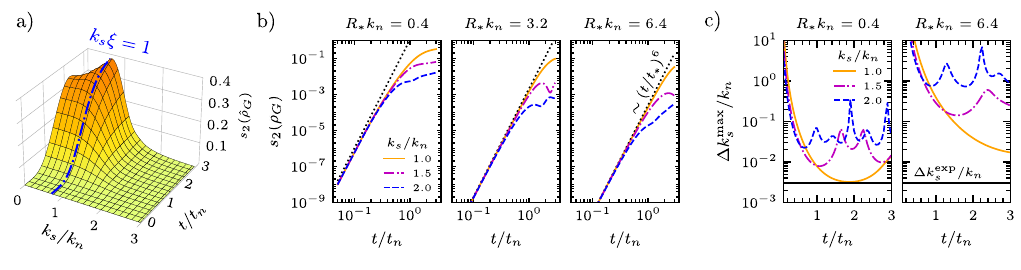}
\caption{\label{fig:ESgrowth} Growth of entanglement entropy in the unitary Bose gas. In Fig.~(a), we plot the normalized second R\'enyi entropy $s_2(\hat{\rho}_G)$ as a function of time and momentum in the beyond-Gaussian model, for $R_* k_n = 0.4$. Blue dashed-dotted line highlights the characteristic shell momentum $k_s \xi = 1$. In Fig.~(b), we highlight the initial power law growth of $s_2(\hat{\rho}_G)$, now at a set of resonance widths ranging from broad to narrow. Black dotted lines show the power law scaling $s_2(\hat{\rho}_G) \sim (t/t_*)^6$ expected to hold at early times $t\ll t_*$. To obtain the dimensionless proportionality constant, we fit this power law to the narrowest resonance in the third panel, and then use the same proportionality constant in the first and second panel. In Fig.~(c), we plot the characteristic shell width $\Delta k_s^{\mathrm{max}}$ that determines the regime of validity for the thin-shell approximation, obtained with the correlation functions computed from our model and the experimental parameters of Ref.~\cite{Eigen2018}. Black horizontal line shows the bin size $\Delta k_s^{\mathrm{exp}}$ reported in Ref.~\cite{Eigen2018} for time-of-flight measurements of the momentum space density.}
\end{figure*}

In Fig.~\ref{fig:Trajectory}, we examine the trajectory of the many-body quantum state through the $(n_{\vb{k}}, \abs{\kappa_{\vb{k}}}^2)$ phase space, for Gaussian and beyond-Gaussian calculations. Following the quench at $t = 0$, both the excitation density $n_{\vb{k}}$ and the pairing amplitude $\abs*{\kappa_{\vb{k}}}$ start to increase due to quantum depletion from the condensate. In the Gaussian model, $\abs*{\kappa_{\vb{k}}}$ grows in such a way that the state stays exactly on the boundary of the allowed and forbidden regions, as defined by Eq.~\eqref{eq:Unc}. Hence, as predicted, this model does not generate entanglement. In contrast, in the beyond-Gaussian model the formation of correlated triples through the processes in the third and fourth panels of Fig.~\ref{fig:GrowthDiag} slow the growth of the pairing amplitude $\abs*{\kappa_{\vb{k}}}$, such that the state departs from the uncertainty boundary and travels into the entangled region of phase space.

Generally, $n_{\vb{k}}$ will keep growing following the quench until the system enters a quasi-equilibrated state, known as the prethermal state \cite{Regemortel2018, Colussi2020}. Here, in the Gaussian model, $n_{\vb{k}}$ will plateau and start to oscillate coherently with a universal Bogoliubov frequency $\omega_{\vb{k}}$, analyzed in detail in Ref.~\cite{Colussi2020}. It defines the associated characteristic healing length $\xi k_n \approx \sqrt{2}$ and prethermalization timescale $\tau_{\vb{k}} \sim 1/\omega_{\vb{k}}$. For $k_s \xi \lesssim 1$, the characteristic excitations behave as long-wavelength phonons ($\omega_{\vb{k}} \sim k$), and the population oscillates slowly with respect to $t_n$. In the beyond-Gaussian model, we find that for these low-momentum modes the pairing amplitude itself reaches a plateau value, after which it stays approximately constant, such that the state gradually travels further into the entangled region \cite{LateTimeRevival}. For larger shell momenta $k_s \xi \gtrsim 1$ we enter the free-particle regime ($\omega_{\vb{k}} \sim k^2$), where the Gaussian model displays fast oscillations along the uncertainty boundary, seen clearly in Fig.~\ref{fig:Trajectory}. Here, in the beyond-Gaussian model, ergodic three-body correlations allow the system to break free from the prethermal state \cite{Regemortel2018, Kraats2024}, such that $n_{\vb{k}}$ increases beyond the maximum value attained in the Gaussian approximation. In this intermediate-time regime, the trajectory of the state through phase space becomes increasingly complicated, as seen particularly in the third and fourth panels of Fig.~\ref{fig:Trajectory}, however we find generally that the departure from the prethermal state is associated with a rapid damping of $\abs*{\kappa_{\vb{k}}}$. 


We now focus on the beyond-Gaussian model, and use Eq.~\eqref{eq:Snk} to calculate the second R\'enyi entropy ($\gamma = 2$) as a function of both shell momentum and time following the quench, plotted in Fig.~\ref{fig:ESgrowth}(a). As we show in the Supplemental Material \cite{SupMat}, our many-body model predicts that at early times $t \ll t_*$, $\tilde{n}_{\vb{k}}$ is independent of momentum and grows according to the power law $\tilde{n}_{\vb{k}} \sim (t/t_*)^6$. Here $t_* = \sqrt{t_n \tau}$ is the geometric mean of the Fermi time $t_n$ and the molecular lifetime $\tau = m R_*/k_n$ \cite{Braun2022}, and can be interpreted as the mean transition time for the processes shown in Fig.~\ref{fig:GrowthDiag}. It follows that, for $\gamma \ge 1$, the R\`enyi entropy obeys the same early time scaling, i.e. $s_{\gamma}(\hat{\rho}) \sim (t /t_*)^6$, and indeed this prediction is confirmed by our numerics as shown for $\gamma = 2$ in Fig.~\ref{fig:ESgrowth}(b). At intermediate times, $t\gtrsim t_*$, where nonlinear effects such as Bose enhancement of the two-body interaction become significant \cite{Kira2015}, the entropy growth is gradually slowed down, and we find that while the entropy for $k_s \xi \lesssim 1$ still varies only weakly with $k_s$, the entropy for $k_s \xi \gtrsim 1$ now displays a rapid decrease with the shell momentum, see Fig.~\ref{fig:ESgrowth}(a). This may be attributed to the universal suppression of excitations at large momenta due to the kinetic energy, which decreases the absolute value of the two point correlation functions $n_{\vb{k}}$ and $\kappa_{\vb{k}}$ \cite{Tan2008, Tan2008_2, Tan2008_3, Werner2012, Werner2012_2}. 


\textit{Validity of thin-shell approximation}--- In all results above we have considered just the lowest order contribution to the entropy with respect to the shell width $\delta k_s$. To examine the validity of this approximation, especially in regards to current experiments, we again employ the coherent-state representation of $\hat{\rho}$ to calculate the lowest order correction to Eq.~\eqref{eq:rhoG}, after which we obtain $\hat{\rho} = \hat{\rho}_G + \hat{\rho}'$, where $\hat{\rho}'$ is a non-Gaussian operator with vanishing trace \cite{SupMat}. Then, we appreciate that the difference between $\hat{\rho}$ and $\hat{\rho}_G$ can be quantified by the \textit{non-Gaussianity} of the state $\hat{\rho}$, for which we use the following measure \cite{Genoni2007, Genoni2008, Baek2018, Park2021},
\begin{equation}
\mathcal{N}(\hat{\rho} || \hat{\rho}_G ) = \frac{1}{2} \frac{\mathrm{Tr} \left[ \left(\hat{\rho}  - \hat{\rho}_G \right)^2 \right]}{\mathrm{Tr}\left[ \hat{\rho}^2 \right]},
\end{equation}
which is a strictly positive function that vanishes if and only if $\hat{\rho}$ is indistinguishable from $\hat{\rho}_G$, in the sense that the Hilbert-Schmidt distance between the two states is zero. As we show in the Supplemental Material \cite{SupMat}, $\mathrm{Tr} \left[ \hat{\rho}_G \hat{\rho}' \right] = 0$, such that the non-Gaussianity of $\hat{\rho}$ scales quadratically with $\hat{\rho}'$. Specifically, we find,
\begin{eqnarray}
&& \mathcal{N}(\hat{\rho} || \hat{\rho}_G ) =\frac{1}{6} \sum_{\vb{k} \in \mathcal{A}, \vb{q} \in \mathcal{A}} \\ && \times \left[ \abs{R_{\vb{k}, \vb{q}}}^2 \mathcal{U}_{\vb{k}, \vb{q}} -  \left(R_{\vb{k},\vb{q}}^*\right)^2 \mathcal{V}_{\vb{k}, \vb{q}} -  \left(R_{\vb{k},\vb{q}}\right)^2 \mathcal{V}_{\vb{k}, \vb{q}}^*\right] \nonumber.
\label{eq:NR}
\end{eqnarray}
Here we have defined,
\begin{equation}
\mathcal{U}_{\vb{k}, \vb{q}} = \prod_{\vb{l}} \frac{\abs{u_{\vb{l}}}^2  + \abs{v_{\vb{l}}}^2}{1 + 2 \tilde{n}_{\vb{l}}}, \quad  \mathcal{V}_{\vb{k}, \vb{q}} = 4\prod_{\vb{l}} \frac{u_{\vb{l}} v_{\vb{l}}^* }{1 + 2\tilde{n}_{\vb{l}}},
\end{equation}
where $\vb{l} \equiv \left\{\vb{k}, \vb{q}, -\vb{k} - \vb{q} \right\}$. Due to the double momentum sum in Eq.~\eqref{eq:NR}, one finds that in the continuum limit $\mathcal{N}(\hat{\rho} || \hat{\rho}_G ) = \bar{\mathcal{N}}(\hat{\rho} || \hat{\rho}_G ) \delta k_s^2$, where $\bar{\mathcal{N}}(\hat{\rho} || \hat{\rho}_G )$ is independent of the shell width. The thin-shell condition $\mathcal{N}(\hat{\rho} || \hat{\rho}_G ) \ll 1$ can then be recast as $\delta k_s \ll \Delta k_s^{\mathrm{max}}$, where $ \Delta k_s^{\mathrm{max}} = \sqrt{1/\bar{\mathcal{N}}(\hat{\rho} || \hat{\rho}_G)}$. Assuming that correlations become independent of the system volume in the thermodynamic limit, we can calculate $\Delta k_s^{\mathrm{max}}$ for a given experiment, and compare with the bin size $\Delta k_s^{\mathrm{exp}}$ for time-of-flight measurements, which determines the effective experimental shell width \cite{Ketterle1999}. As an example, we use the quench experiment of Ref.~\cite{Eigen2018}, which reports a bin size $\Delta k_s^{\mathrm{exp}} \approx 0.02 \ \mathrm{\mu m}^{-1}$ with respect to a total system volume $V = 3 \cdot 10^4 \ \mathrm{\mu m}^{3}$ and Fermi momentum $k_n = 6.7 \ \mathrm{\mu m}^{-1}$. In Fig.~\ref{fig:ESgrowth}(c) we show the resulting value of $\Delta k_s^{\mathrm{max}}$. Comparing with the bin size, we observe that the thin-shell approximation for this experiment is valid for the majority of momenta, but an improvement in resolution is desirable near the Fermi momentum at later times, where $R_{\vb{k},\vb{q}}$ is largest. We also observe that for narrower Feshbach resonances, the slower generation of beyond-Gaussian correlations as predicted in Refs. \cite{Braun2022, Kraats2024}, extends the validity of the thin-shell approximation, such that the width of the Feshbach resonance can be used as an additional control on the method. Hence, measuring thin-shell entanglement clearly presents both an opportunity and challenge for future experiments.

It is important to note that we have assumed that the initial many-body state is pure, which is only an approximation for ultracold experiments. Although theoretically more involved, notions of bipartite entanglement also exist for mixed states \cite{Amico2008}, motivating an extension of our method in this direction. Going forward, our framework can be applied to study other quantum gases where non-Gaussian correlations are expected to be significant, including superfluid helium \cite{Dmowski2017, Tenart2021, Bureik2025}, multicomponent Fermi gases \cite{Navon2010, Mukherjee2019, Ji2022, Schumacher2023}, and Bose-Fermi mixtures \cite{Patel2023, Shen2024, Kraats2024_BF, Chuang2024}.  Our work also creates opportunities for further exploration of quantum resources in continuous variable systems, such as the entanglement spectrum \cite{Li2008} and quantum Fisher information \cite{Strobel2014, Hauke2016, Benavides2024}.

\begin{acknowledgments}
We thank M. Wouters, R. Santos, J. Postema and R. de Keijzer for fruitful discussions. J.v.d.K. and S.J.J.M.F.K. acknowledge financial support from the Netherlands Organisation for Scientific research (NWO) under Grant No. 680.92.18.05, and from the Dutch Ministry of Economic Affairs and Climate Policy (EZK), as part of the Quantum Delta NL program. D.J.M.A.B. acknowledges funding by the Research
 Foundation–Flanders via a postdoctoral fellowship (Grant No. 1222425N).
\end{acknowledgments}

\bibliography{References}

\end{document}


\title{Supplementary Material: ``Probing non-Gaussian correlations through entanglement generation in a many-body quantum system"}

\author{J. van de Kraats}
\email{j.v.d.kraats@tue.nl}
\affiliation{Department of Applied Physics and Science Education, Eindhoven University of Technology, P. O. Box 513, 5600 MB Eindhoven, The Netherlands}
\author{D.J.M. Ahmed-Braun}
\affiliation{TQC, Departement Fysica, Universiteit Antwerpen, Universiteitsplein 1, B-2610 Antwerpen, Belgium}
\author{V.E. Colussi}
\affiliation{Quantinuum, 303 S Technology Ct, Broomfield, CO 80021, USA}
\author{S.J.J.M.F. Kokkelmans}
\affiliation{Department of Applied Physics and Science Education, Eindhoven University of Technology, P. O. Box 513, 5600 MB Eindhoven, The Netherlands}
\date{\today}

\begin{abstract}
\end{abstract}
\maketitle

\tableofcontents

\section{Introduction}

This supplemental material provides additional detail and derivations that support the claims in the main text. In Sec.~\ref{sec:model} we introduce the model that we use to simulate the quench dynamics of the unitary Bose gas numerically, introducing also the cumulant expansion which stands at the center of our approach. In Sec.~\ref{sec:Gauss}, we show how the entanglement entropy can be calculated for a general Gaussian state by explicitly diagonalizing the reduced density matrix, leading us to introduce the entanglement eigenmodes and associated occupation numbers. In Sec.~\ref{sec:growth}, we derive analytically how the eigenmode occupation, and by extension the entanglement entropy, grows as a function of the time following the quench, recovering the power-law discussed in the main text. Finally, in Sec.~\ref{sec:NGgen}, we extend the analysis of Sec.~\ref{sec:Gauss} to a non-Gaussian state, which allows us to calculate the first order correction to the Gaussian density matrix that characterizes the thin-shell approximation, and obtain the non-Gaussianity measure discussed in the main text.

\section{Cumulant model of many-body dynamics}
\label{sec:model}

In this section we outline the many-body model used for calculating the two-body and three-body correlation functions in Eq. (1) of the main text, formulated originally in Ref.~\cite{Kraats2024}, and based on the models in Refs.~\cite{Kira2015, Kira2015_2, Colussi2020, Braun2022}. We consider a 3-dimensional many-body system of $N$ identical bosons of mass $m$ in quantization volume $V$, with density $n = N/V$, interacting near a magnetic Feshbach resonance. The system is described by the microscopic two-channel Hamiltonian \cite{Timmermans1999, Kokkelmans2002, Kokkelmans2002_2, Gurarie2007},
\begin{align}
\hat{H} = \sum_{\vb{k}} \frac{  k^2}{2m}  \hat{a}_{\vb{k}}^{\dagger} \hat{a}_{\vb{k}} + \sum_{\vb{k}} \left( \frac{ k^2}{4m}  + \nu\right) \hat{m}_{\vb{k}}^{\dagger} \hat{m}_{\vb{k}} + \frac{g}{2\sqrt{V}}\sum_{\vb{k}, \vb{q}} \left[\zeta(\vb{k} + 2\vb{q}) \  \hat{m}_{\vb{k}}^{\dagger} \hat{a}_{\vb{k}+\vb{q}} \hat{a}_{-\vb{q}}  + \mathrm{H.c.} \right].
\label{eq:H}
\end{align}
Here the operators $\hat{a}_{\vb{k}}(\hat{m}_{\vb{k}})$ annihilate atoms(molecules) with momentum $\vb{k}$, $\abs*{\vb{k}} \equiv k$, and we use units where $\hbar = 1$. The molecules have internal energy $\nu$ relative to the atomic scattering threshold. The Feshbach interaction, with strength $g$ and momentum-space form factor $\zeta(2\vb{q})$, allows two atoms to associate into a molecule and vice-versa. We define $\zeta(2\vb{q}) = \Theta(\Lambda - q)$, where $\Theta$ is the Heaviside step function such that $\Lambda$ defines a cutoff on the relative two-body momentum, regularizing the UV divergence characteristic of a contact potential \cite{Kokkelmans2002}. The renormalization relations \cite{Kokkelmans2002, Braun2022},
\begin{align}
g^2 = \frac{8 \pi }{m^2 R_*}, \qquad \nu = \frac{1}{m R_*} \left(\frac{2 \Lambda}{\pi} - \frac{1}{a} \right)
\end{align}
relate the Feshbach resonance parameters $a$ and $R_*$ to the microscopic parameters of the model. Note that in the Hamiltonian $\hat{H}$ interactions between atoms are always mediated by closed-channel molecules, which implies a Feshbach resonance with vanishing background scattering length \cite{Chin2010}. This assumption, which is expected to be accurate close to resonance, significantly simplifies the resulting many-body equations \cite{Kraats2024}. 

We work in the U(1) symmetry breaking picture, where the $\vb{k} = 0$ modes in both the atomic and molecular species form Bose-Einstein condensates with associated wave functions $\expval*{\hat{a}_{\vb{0}}} = \sqrt{V} \psi$ and $\expval*{\hat{m}_{\vb{0}}} = \sqrt{V} \phi$ \cite{Pitaevskii2016}. To model the quench we assume that for $t<0$ all particles are in the atomic BEC, such that $\abs{\psi}^2 = n$. Then, at $t =0$, we instantaneously ramp the detuning $\nu$ to its resonant value where $1/a = 0$, initiating the quench. The subsequent evolution of the expectation value $\expval*{\hat{O}}$ of some arbitrary many-body operator $\hat{O}$ is described by the Heisenberg equation,
\begin{align}
i \pdv{t} \expval*{\hat{O}} = \expval*{[\hat{O}, \hat{H}]}.
\label{eq:HeisEOM}
\end{align}
For a study of the correlation dynamics in the gas, it is useful to decompose $\expval*{\hat{O}}$ in connected correlation functions known as ``cumulants", or ``clusters" \cite{Fricke1996, Kohler2002, Kira2012}. If we assume $\hat{O}$ can be written as a normal ordered product of single-particle operators, then it is uniquely defined by two $2M$-dimensional arrays of non-negative integers $\bar{\vb{n}} = \left\{\bar{n}_i \right\}_{i = 1, ..., 2M}$ and $\vb{n} = \left\{n_i \right\}_{i = 1, ..., 2M}$ as,
\begin{align}
\hat{O} = \prod_{i = 1}^{2M} (\hat{c}_{\vb{k}_i}^{\dagger})^{\bar{n}_i} \prod_{j = 1}^{2M} (\hat{c}_{\vb{q}_j})^{n_j},
\end{align}
where $M$ is the total number of momentum modes, and $\hat{c}_{\vb{k}_i} = \hat{a}_{\vb{k}_i}$ for $1 \le i \le M$, $\hat{c}_{\vb{k}_i} = \hat{m}_{\vb{k}_i}$ for $M+1 \le i \le 2M$. Then, the cumulant $\expval*{\hat{O}}_{\mathrm{c}}$ is defined by the generating function \cite{Kohler2002},
\begin{align}
\expval*{\hat{O}}_{\mathrm{c}} = \expval*{\prod_{i = 1}^{2M} (\hat{c}_{\vb{k}_i}^{\dagger})^{\bar{n}_i} \prod_{j = 1}^{2M} (\hat{c}_{\vb{q}_j})^{n_j}}_{\mathrm{c}} =  \prod_{i = 1}^{2M} \left(\pdv{x_{i}}\right)^{\bar{n}_i} \prod_{j = 1}^{2M}  \left(\pdv{y_{j}^*} \right)^{n_j}  \ln \expval{e^{\sum_{i = 1}^{2M} x_i\hat{c}_{\vb{k}_i}^{\dagger}} e^{\sum_{j = 1}^{2M} y_j^*\hat{c}_{\vb{q}_j}}} \bigg|_{\vb{x}, \vb{y} = 0}.
\label{eq:CumDef}
\end{align}
Due to the classical nature of the condensates, cumulants of order greater than 1 that contain the condensate mode $\vb{k} = 0$ automatically vanish, such that the cumulants of second and third order operator products are simply equal to the associated expectation values. Additionally, translational invariance dictates that cumulants must conserve momentum, such that $\sum_{i} \bar{n}_i \vb{k}_i - \sum_j n_j \vb{q}_j = 0$. 

Expressed in terms of cumulants, Eq.~\eqref{eq:HeisEOM} expands into a set of coupled nonlinear first-order differential equations. Since the complete set of equations is intractable for a many-body system, the expansion must be truncated to a specific set of lower order cumulants. In the quench scenario, higher order cumulants grow sequentially in time, as more and more excitations are generated from the BEC \cite{Kira2015, Kira2015_2, Colussi2020}. Hence a truncation of the cumulant expansion should be interpreted as an early time approximation to the post-quench correlation dynamics. In the lowest order mean-field approximation, one retains only the condensate fields $\psi$ and $\phi$, whose dynamics are governed by coupled Gross-Pitaevskii equations. While this model is very simple, it does not contain the excitations central to the quench dynamics, and also can not reproduce the vacuum two-body physics characteristic of the Feshbach resonance, such as a weakly bound dimer state \cite{Kokkelmans2002_2}. To fix these shortcomings, the Hartree-Fock-Bogoliubov (HFB) approximation introduces the Gaussian second order cumulants, or doublets \cite{Blazoit1985},
\begin{align}
n_{\vb{k}} = \expval*{\hat{a}_{\vb{k}}^{\dagger} \hat{a}_{\vb{k}}}_{\mathrm{c}}, \qquad \kappa_{\vb{k}} = \expval*{\hat{a}_{\vb{k}} \hat{a}_{-\vb{k}}}_{\mathrm{c}}.
\label{eq:atomicdoublets}
\end{align}
The resulting model is appropriate for describing the early time dynamics following the dynamics, including the influence of the width of the Feshbach resonance, as analysed in detail in Ref.~\cite{Braun2022}. The scattering processes that drive this dynamics are shown schematically in the first two panels of Fig. 1(a) in the main text. As shown in Ref.~\cite{Colussi2020}, following an initial growth of excitations the HFB approximation ends up in a universal quasi-equilibrium state, also referred to as the prethermal state, which is a signature of its integrable nature \cite{Regemortel2018}. Indeed, the HFB approximation follows naturally when one assumes the many-body quantum state to be Gaussian, and applying Wick's theorem to decompose the quantum correlations \cite{Fetter2003}.

As already pointed out in Ref.~\cite{Colussi2020}, the post-quench dynamics following the formation of the universal prethermal stage can no longer be modelled accurately by the HFB or Gaussian approximation, due to the significant influence of higher order non-Gaussian cumulants. In Ref.~\cite{Kraats2024}, it was shown that, with the Hamiltonian in Eq.~\eqref{eq:H}, the dynamics at these intermediate times can be reproduced remarkably well by formulating a beyond-HFB or beyond-Gaussian model, which includes the additional mixed atom-molecule doublet cumulants,
\begin{eqnarray}
 n_{\vb{k}}^m &= \expval*{\hat{m}_{\vb{k}}^{\dagger} \hat{m}_{\vb{k}}}_{\mathrm{c}}, \qquad \kappa_{\vb{k}}^m = \expval*{\hat{m}_{\vb{k}} \hat{m}_{-\vb{k}}}_{\mathrm{c}}, \label{eq:mixeddoublets}  \\
\chi_{\vb{k}}&= \expval*{\hat{m}_{\vb{k}}^{\dagger} \hat{a}_{\vb{k}}}_{\mathrm{c}}, \qquad\kappa_{\vb{k}}^{am} = \expval*{\hat{m}_{\vb{k}} \hat{a}_{-\vb{k}}}_{\mathrm{c}}, \nonumber
\end{eqnarray}
and the non-Gaussian triplet cumulant,
\begin{eqnarray}
R_{\vb{k}, \vb{q}} = \expval*{\hat{a}_{\vb{k}} \hat{a}_{\vb{q}} \hat{a}_{-\vb{k} -\vb{q}}}_{\mathrm{c}},
\label{eq:triplets}
\end{eqnarray}
also introduced in the main text. In essence, this model extends the Gaussian approximation by including molecules at finite momentum, which by virtue of the last two panels in Fig. 1(a) in the main text, are intimately connected to the generation of non-Gaussian three-body correlations, as quantified by $R_{\vb{k}, \vb{q}}$.

The full equations of motion for the beyond-Gaussian model read as follows. First we have the coupled Gross-Pitaevskii equations,
\begin{align}
i  \pdv{t}\psi &= g \left(\zeta(0) \phi \psi^* + \frac{1}{V}\sum_{\vb{q}}  \zeta(\vb{q}) \chi_{\vb{q}}^* \right), \label{eq:psia} \\
i   \pdv{t}\phi &= \nu \phi +  \frac{g}{2} \left(\zeta(0) \psi^2 + \frac{1}{V}\sum_{\vb{q}} \zeta(2\vb{q}) \kappa_{\vb{q}} \right). \label{eq:psim}
\end{align}
Then, the atomic doublet correlations in Eq.~\eqref{eq:atomicdoublets} obey,
\begin{align}
 \pdv{t}n_{\vb{k}} &= 2 g \zeta(2\vb{k}) \mathrm{Im} \left(\phi \kappa_{\vb{k}}^{*}  \right)  - 2 g \zeta(\vb{k}) \mathrm{Im} \left(\psi \chi_{\vb{k}} \right), \label{eq:nka}\\
i   \pdv{t}\kappa_{\vb{k}} &= 2 \varepsilon_{\vb{k}} \kappa_{\vb{k}} + g\zeta(2\vb{k})  (1 +  2n_{\vb{k}} ) \phi   +  2g \zeta(\vb{k}) \kappa_{\vb{k}}^{am}  \psi^* \label{eq:kappaa}.
\end{align}
Here $\varepsilon_{\vb{k}} =  k^2/(2m)$ gives the atomic kinetic energy. In the beyond-Gaussian model, the additional cumulants in Eqs.~\eqref{eq:mixeddoublets} and \eqref{eq:triplets} evolve as,
\begin{align}
i   \pdv{t}\kappa_{\vb{k}}^{am} &= \left(\varepsilon_{\vb{k}} + \varepsilon_{\vb{k}}^m \right)\kappa_{\vb{k}}^{am} +g \zeta(\vb{k})\left( \kappa_{\vb{k}} \psi +  \kappa_{\vb{k}}^m\psi^* \right) + g \zeta(2\vb{k}) \chi_{\vb{k}}^* \phi  + \frac{g}{2\sqrt{V}} \sum_{\vb{q}} \zeta( 2\vb{q} + \vb{k}) R_{\vb{k},\vb{q}}^a,  \\
i   \pdv{t} \chi_{\vb{k}} &= \left(\varepsilon_{\vb{k}} - \varepsilon_{\vb{k}}^m \right) \chi_{\vb{k}} - g \zeta(\vb{k})  \left(n_{\vb{k}} - n_{\vb{k}}^m \right) \psi^* + g\zeta(2\vb{k}) \kappa_{\vb{k}}^{am*} \phi, \label{eq:chi} \\
 \pdv{t}n_{\vb{k}}^m &= 2 g \zeta(\vb{k}) \mathrm{Im}\left(\psi \chi_{\vb{k}}\right), \\
i   \pdv{t}\kappa_{\vb{k}}^m &= 2 \varepsilon_{\vb{k}}^m  \kappa_{\vb{k}}^m + 2 g \zeta(\vb{k}) \kappa_{\vb{k}}^{am} \psi, \\
i   \pdv{t}R_{\vb{k},\vb{q}} &= \mathcal{\hat{S}}\bigg\{ \varepsilon_{\vb{k}}  R_{\vb{k},\vb{q}}^a + \frac{g}{\sqrt{V}}(1 + n_{\vb{k}} + n_{\vb{q}}) \zeta(\vb{k} - \vb{q}) \kappa_{\vb{k} + \vb{q}}^{am} \bigg\} \label{eq:R}
\end{align}
Here $\varepsilon_{\vb{k}}^m =  k^2/(4m) + \nu$ gives the kinetic energy of the molecules, and $\mathcal{\hat{S}}$ is a three-body symmetrization operator defined as $\mathcal{\hat{S}} = 1 + \hat{P}_+ + \hat{P}_-$, where $\hat{P}_{+(-)}$ define (anti)symmetric cyclic permutation operators of the three particle indices. As noted in Refs.~\cite{Kira2015, Kraats2024}, Eqs.~\eqref{eq:kappaa} and \eqref{eq:R} contain the so called Bose enhancement factors $(1+2 n_{\vb{k}})$ and $(1+ n_{\vb{k}} + n_{\vb{q}})$, which represent the nonlinear enhancement of the two-body interaction when scattering to modes that are already occupied by bosons. The inclusion of these nonlinearities is crucial for obtaining a correct model of the many-body dynamics.

By integrating the coupled system of equations numerically, we obtain all cumulants as a function of time $t$ following the quench. In this calculation we assume the thermodynamic limit where $V \rightarrow \infty$, such that all momentum sums are converted into 3D integrals. Taking into account spherical symmetry, this means that doublet cumulants become 1-dimensional vectors, e.g. $\kappa_{\vb{k}} \rightarrow \kappa(k)$. Similarly, triplet cumulants become 3-dimensional tensors, $R_{\vb{k},\vb{q}} \rightarrow R(k, q, \theta)$, where $\theta$ is the angle between $\vb{k}$ and $\vb{q}$. In all our simulations we set $\Lambda/n^{1/3} = 60$. Given the relation $\Lambda = 2/(\pi \bar{a})$ between the cutoff and the characteristic two-body interaction length scale $\bar{a} = 0.955978 \ r_{\mathrm{vdW}}$, where $r_{\mathrm{vdW}}$ is the van der Waals length \cite{Colussi2020, Kraats2024}, this corresponds to a gas density $n r_{\mathrm{vdW}}^3 = 1.37 \cdot 10^{-6}$. For more details regarding the numerical implementation, see Ref.~\cite{Colussi2020}.

\section{Entanglement entropy for Gaussian states}
\label{sec:Gauss}

We now consider the calculation of the reduced density matrix for thin-shell bipartitions of the atomic momentum space, as discussed in the main text, and the associated entanglement entropy. Within the shell space $\mathcal{A}$, the relevant correlations are $n_{\vb{k}}, \kappa_{\vb{k}}$ and $R_{\vb{k}, \vb{q}}$, of which $R_{\vb{k}, \vb{q}}$ is non-Gaussian. Before considering this full set, it is instructive to restrict ourselves initially to the thin-shell approximation introduced in the main text, where it is assumed that $R_{\vb{k}, \vb{q}}$ predominantly affects the thin-shell density matrix through its back action on the doublets $n_{\vb{k}}$ and $\kappa_{\vb{k}}$. Then, in Sec.~\eqref{sec:NGgen}, we will motivate this assumption by an extended calculation that also includes the direct effects of $R_{\vb{k}, \vb{q}}$ to the lowest order in the shell width. For convenience, we will use the following notation for intrashell momentum sums/products,
\begin{align}
\sum_{\vb{k} \in \mathcal{A}} \equiv \overline{\sum_{\vb{k}}}, \qquad \prod_{\vb{k} \in \mathcal{A}} \equiv \overline{\prod_{\vb{k}}}.
\end{align}

Our treatment follows Ref.~\cite{Frerot2015}, specified to the bosonic case. We define $M_{\mathcal{A}}$ as the number of modes in the thin-shell space $\mathcal{A}$. As noted we consider here the case where the reduced density matrix is Gaussian, leaving the generalization to non-Gaussian states for Sec.~\ref{sec:NGgen}. Then, with two-point correlation functions $n_{\vb{k}}$ and $\kappa_{\vb{k}}$, the entanglement Hamiltonian has the form,
\begin{align}
\mathcal{H} = \overline{\sum_{\vb{k} > 0 }} \left[ A_{\vb{k}} \hat{a}_{\vb{k}}^{\dagger} \hat{a}_{\vb{k}} + A_{\vb{k}} \hat{a}_{-\vb{k}}^{\dagger} \hat{a}_{-\vb{k}} + B_{\vb{k}}  \hat{a}_{\vb{k}}^{\dagger} \hat{a}_{-\vb{k}}^{\dagger} + B_{\vb{k}}^*  \hat{a}_{\vb{k}} \hat{a}_{-\vb{k}} \right],
\label{eq:Hent}
\end{align}
where $\vb{k} > 0$ means that we sum over one half of the shell $\mathcal{A}$, and $A_{\vb{k}}(B_{\vb{k}})$ are $\frac{1}{2} M_{\mathcal{A}}$ real(complex) numbers. In matrix notation, Eq.~\eqref{eq:Hent} can be rewritten as,
\begin{align}
\mathcal{H} =  \frac{1}{2} \overline{\sum_{\vb{k} > 0 }}  \hat{\vb*{\alpha}}_{\vb{k}}^{\dagger} \Omega \mathcal{L}_{\vb{k}} \hat{\vb*{\alpha}}_{\vb{k}} -   \overline{\sum_{\vb{k} > 0 }} A_{\vb{k}},
\end{align}
where we have defined,
\begin{align}
\hat{\vb*{\alpha}}_{\vb{k}} = \begin{pmatrix} 
    \hat{a}_{\vb{k}} \\ \hat{a}_{-\vb{k}} \\ \hat{a}_{\vb{k}}^{\dagger}  \\  \hat{a}_{-\vb{k}}^{\dagger} 
\end{pmatrix}, \qquad \Omega = 
\begin{pmatrix} 
    1 & 0 & 0 & 0 \\
    0 & 1 & 0 & 0 \\
    0 & 0 & -1 & 0 \\
    0 & 0 & 0 & -1 
\end{pmatrix} \qquad \mathcal{L}_{\vb{k}} = 
\begin{pmatrix}
    A_{\vb{k}} & 0 & 0 & B_{\vb{k}} \\
    0 & A_{\vb{k}} & B_{\vb{k}} & 0 \\
    0 & -B_{\vb{k}}^* & -A_{\vb{k}} & 0 \\
    -B_{\vb{k}}^* & 0 & 0 & -A_{\vb{k}} 
\end{pmatrix},
\label{eq:HentMatDef}
\end{align}
This 4-vector notation, which we will use throughout this supplemental material, treats the $\pm \vb{k}$ modes on equal footing, thus simplifying many of the coming derivations. 

We now introduce the spectral decomposition $\mathcal{L}_{\vb{k}} = U_{\vb{k}} D_{\vb{k}} U_{\vb{k}}^{-1}$, and define a transformed vector of bosonic single-particle operators $\hat{\vb*{\beta}}_{\vb{k}}$ by $\hat{\vb*{\alpha}}_{\vb{k}} = U_{\vb{k}} \vb*{\beta}_{\vb{k}}$. One can show that this transformation has the following properties \cite{Frerot2015},
\begin{align}
\hat{\vb*{\beta}}_{\vb{k}} = \begin{pmatrix} 
    \hat{b}_{\vb{k}} \\ \hat{b}_{-\vb{k}} \\ \hat{b}_{\vb{k}}^{\dagger} \\ \hat{b}_{-\vb{k}}^{\dagger}
\end{pmatrix}, \qquad
D_{\vb{k}} = \begin{pmatrix} 
    \lambda_{\vb{k}} & 0 & 0 & 0\\
    0 & \lambda_{-\vb{k}} & 0 & 0\\
    0 & 0 & -\lambda_{\vb{k}} & 0\\
    0 & 0  & 0 & -\lambda_{-\vb{k}}\\
\end{pmatrix}, \qquad U_{\vb{k}}^{\dagger} = \Omega U_{\vb{k}}^{-1} \Omega,
\label{eq:BogTransId}
\end{align}
where we have introduced the eigenmodes $\hat{b}_{\vb{k}}, \hat{b}_{-\vb{k}}$ with associated eigenvalues $\lambda_{\vb{k}}, \lambda_{-\vb{k}}$, and we note that, due to spherical symmetry, $\lambda_{\vb{k}} = \lambda_{-\vb{k}}$. The third identity in Eq.~\eqref{eq:BogTransId} indicates that $U_{\vb{k}}$ is a symplectic transformation, which means it preserves the bosonic commutation relations \cite{Blazoit1985}.  The entanglement Hamiltonian now attains the diagonal form,
\begin{align}
\mathcal{H} = \frac{1}{2} \overline{\sum_{\vb{k}>0}} \hat{\vb*{\beta}}_{\vb{k}}^{\dagger} \Omega D_{\vb{k}} \hat{\vb*{\beta}}_{\vb{k}} - \overline{\sum_{\vb{k}>0}} A_{\vb{k}} = \overline{\sum_{\vb{k}}} \lambda_{\vb{k}} \hat{b}_{\vb{k}}^{\dagger} \hat{b}_{\vb{k}} + \frac{1}{2} \overline{\sum_{\vb{k}}} \lambda_{\vb{k}} - \frac{1}{2} \overline{\sum_{\vb{k}}} A_{\vb{k}} .
\end{align}
Consider now the correlation matrix $C_{\vb{k}}$,
\begin{align}
C_{\vb{k}} &= \expval*{\hat{\vb*{\alpha}}_{\vb{k}} \hat{\vb*{\alpha}}_{\vb{k}}^{\dagger} } =  U_{\vb{k}}\expval*{\hat{\vb*{\beta}}_{\vb{k}} \hat{\vb*{\beta}}_{\vb{k}}^{\dagger} }U_{\vb{k}}^{\dagger}= U_{\vb{k}}\begin{pmatrix} 
    1 + \tilde{n}_{\vb{k}} & 0 & 0 & 0\\
    0 & 1 + \tilde{n}_{-\vb{k}} & 0 & 0\\
    0 & 0 &  \tilde{n}_{\vb{k}} & 0\\
    0 & 0  & 0 &  \tilde{n}_{-\vb{k}}\\
\end{pmatrix} U_{\vb{k}}^{\dagger},
\label{eq:Cdef}
\end{align}
where the eigenmode occupation numbers follow from the Bose-Einstein distribution function as $\tilde{n}_{\vb{k}} = \mathrm{Tr}[\hat{\rho} \hat{b}_{\vb{k}}^{\dagger} \hat{b}_{\vb{k}} ] = 1/(e^{\lambda_{\vb{k}}} - 1)$. We can rewrite as,
\begin{align}
-C_{\vb{k}} \Omega &=  U_{\vb{k}}\begin{pmatrix} 
    -1 - \tilde{n}_{\vb{k}} & 0 & 0 & 0\\
    0 & -1 - \tilde{n}_{-\vb{k}} & 0 & 0\\
    0 & 0 &  \tilde{n}_{\vb{k}} & 0\\
    0 & 0  & 0 &  \tilde{n}_{-\vb{k}}\\
\end{pmatrix} U_{\vb{k}}^{-1},
\end{align}
showing that upon diagonalizing $-C_{\vb{k}} \Omega$ for all $\vb{k} \in \mathcal{A}, \vb{k} > 0$, we directly obtain the $M_{\mathcal{A}}$ mode occupations $\tilde{n}_{\vb{k}}, \tilde{n}_{-\vb{k}}$. In terms of doublet cumulants, $-C_{\vb{k}} \Omega$ reads,
\begin{align}
-C_{\vb{k}} \Omega = \begin{pmatrix} 
-1 - n_{\vb{k}} & 0 & 0 & \kappa_{\vb{k}} \\
0 & -1 - n_{\vb{k}} & \kappa_{\vb{k}} & 0 \\
0 & -\kappa_{\vb{k}}^* & n_{\vb{k}} & 0 \\
- \kappa_{\vb{k}}^* & 0 & 0 & n_{\vb{k}},
\end{pmatrix}
\end{align}
which leads to the expression for $\tilde{n}_{\vb{k}}$ in Eq. (4) of the main text, where we note again the spherical symmetry $\tilde{n}_{\vb{k}} =  \tilde{n}_{-\vb{k}}$. 

For completeness we also note the transformation matrices $U_{\vb{k}}$, which read,
\begin{align}
\begin{split}
U_{\vb{k}} = \begin{pmatrix}
u_{\vb{k}} & 0  & 0 &  v_{\vb{k}}^* \\
0 &  u_{\vb{k}}  & v_{\vb{k}}^* &  0 \\
0 & v_{\vb{k}}  & u_{\vb{k}}^*  & 0 \\
v_{\vb{k}} & 0  & 0  &  u_{\vb{k}}^*
\end{pmatrix} \qquad 
U_{\vb{k}}^{-1} = \begin{pmatrix}
u_{\vb{k}}^* & 0  & 0 &  -v_{\vb{k}}^* \\
0 &  u_{\vb{k}}^*  & -v_{\vb{k}}^* &  0 \\
0 & -v_{\vb{k}}  & u_{\vb{k}}  & 0 \\
-v_{\vb{k}} & 0  & 0  &  u_{\vb{k}}
\end{pmatrix}
\end{split}
\label{eq:Uk}
\end{align}
where,
\begin{align}
\begin{split}
u_{\vb{k}} &= \frac{1 + 2 n_{\vb{k}} + \sqrt{(1 + 2 n_{\vb{k}})^2 - 4 \abs{\kappa_{\vb{k}}}^2}}{\sqrt{\left(1 + 2 n_{\vb{k}} + \sqrt{(1 + 2 n_{\vb{k}})^2 - 4 \abs{\kappa_{\vb{k}}}^2}\right)^2 - 4 \abs{\kappa_{\vb{k}}}^2}} \\
v_{\vb{k}} &= \frac{2 \kappa_{\vb{k}}^*}{\sqrt{\left(1 + 2 n_{\vb{k}} + \sqrt{(1 + 2 n_{\vb{k}})^2 - 4 \abs{\kappa_{\vb{k}}}^2}\right)^2 - 4 \abs{\kappa_{\vb{k}}}^2}}.
\end{split}
\end{align}
We can also write down the reduced density matrix,
\begin{align}
\begin{split}
\hat{\rho}_G = \frac{1}{\overline{\prod}_{\vb{k}} (1 + \tilde{n}_{\vb{k}})} e^{- \overline{\sum}_{\vb{k}} \lambda_{\vb{k}} \hat{b}_{\vb{k}}^{\dagger} \hat{b}_{\vb{k}}},
\end{split}
\label{eq:rhoG}
\end{align}
which obeys the required normalization $\mathrm{Tr} [\hat{\rho}_G] = 1$.

\section{Early time growth of entanglement entropy}
\label{sec:growth}

Here we consider the initial growth of the entanglement entropy, by analyzing the early time limit of the cumulant equations of motion. First, it was derived in the main text that the Robertson uncertainty relation asserts that (Eq.~(6)),
\begin{align}
x_{\vb{k}} \equiv (1 + 2 n_{\vb{k}})^2 - 4 \abs*{\kappa_{\vb{k}}}^2 \ge 1,
\end{align}
and that a vanishing entropy corresponds with saturation of the uncertainty relation, i.e. $x_{\vb{k}} = 1$. We note that in Ref.~\cite{Kira2015}, the minimum uncertainty condition $x_{\vb{k}} = 1$ was enforced to derive approximations to the triplet dynamics. Such an approximation assumes by construction that the entanglement entropy is zero at all times, and is thus insufficient for our studies, where we explicitly simulate the triplet. Using Eqs.~\eqref{eq:nka} and \eqref{eq:kappaa} we can write down the time derivative of $x_{\vb{k}}$,
\begin{align}
\pdv{t}x_{\vb{k}} = 8 g \zeta(\vb{k}) \left[2 \mathrm{Im} (\kappa_{\vb{k}} \psi \kappa_{\vb{k}}^{am*}) - (1 + 2 n_{\vb{k}}) \mathrm{Im}(\psi \chi_{\vb{k}}) \right].
\end{align}
Evidently, in the Gaussian approximation where $\kappa_{\vb{k}}^{\mathrm{am}} = \chi_{\vb{k}} = 0$, $\dot{x}_{\vb{k}} = 0$ and the uncertainty relation remains saturated at all times. In the beyond-Gaussian model, $x_{\vb{k}}$ departs from the uncertainty boundary due to two distinct source terms. As pointed out in Ref.~\cite{Kraats2024}, the second term, scaling with $\chi_{\vb{k}}$, is a non-universal source whose impact on low momentum modes $k \ll \Lambda$ is generally negligible, unless the underlying Feshbach resonance is very narrow \cite{Wang2024}. In contrast, the first term, scaling with $\kappa_{\vb{k}}^{am}$, is a true three-body correlation significant also for broad resonances. In fact, upon taking the vacuum or short-range limit of the coupled equations of motion for $R_{\vb{k},\vb{q}}$ and $\kappa_{\vb{k}}^{am}$ one recovers the three-body Schr\"odinger equation, where $R_{\vb{k},\vb{q}}$ and $\kappa_{\vb{k}}^{am}$ respectively act as the open and closed-channel components of the three-body wave function \cite{Kraats2024}.

To understand the early time dynamics of $x_{\vb{k}}$, we follow Ref.~\cite{Braun2022} by assuming that, at early times, the atomic condensate stays approximately constant at $\psi = \sqrt{n}$, while all other cumulants are zero. Then, the cumulant equations are coupled through an effective interaction energy $g \sqrt{n}$, whose inverse defines a characteristic timescale,
\begin{align}
g \sqrt{n} = \sqrt{\frac{8}{3 \pi}} \frac{1}{t_*},
\end{align}
Here $t_* = \sqrt{\tau t_n}$ is the geometric mean of the density time scale $t_n$ and the molecular lifetime $\tau = m R_*/k_n$. 

Looking at the cumulant equations, we recognize a hierarchical structure where the atomic condensate first sources the growth of the molecular condensate, which in turn sources the growth of $\kappa_{\vb{k}}$, which subsequently sources $\kappa_{\vb{k}}^{am}$ and $\chi_{\vb{k}}$. The first three panels of Fig. 1 in the main text in fact give a schematic representation of this hierarchical coupling for the growth of $\kappa_{\vb{k}}^{am}$. We can use this hierarchy to integrate the equations step by step, assuming any homogeneous contributions due to kinetic energies are negligible at early times. Then we find the following (momentum independent) time scalings,
\begin{align}
\psi \sim \left(\frac{t}{t_*} \right)^0 \sqrt{n}, \quad \phi \sim  \left(\frac{t}{t_*} \right)^1 \sqrt{n}, \quad \kappa_{\vb{k}} \sim \left(\frac{t}{t_*} \right)^2, \quad \kappa_{\vb{k}}^{am} \sim \left(\frac{t}{t_*}\right)^3, \quad \chi_{\vb{k}} \sim \left(\frac{t}{t_*}\right)^5,
\end{align}
where the proportionality constant is dimensionless. Substituting back into the differential equation for $x_{\vb{k}}$ and integrating once more then gives,
\begin{align}
x_{\vb{k}} - 1  \underset{t/t_* \ll 1}{\sim}  \left(\frac{t}{t_*} \right)^6,
\end{align}
This in turn implies that $\tilde{n}_{\vb{k}} \sim (t/t_*)^6$, which may be substituted into the R\'enyi entropy, after which one obtains,
\begin{equation}
S_{\gamma}(\hat{\rho}_G) \sim \left(\frac{t}{t_*} \right)^6,
\label{eq:Snk}
\end{equation}
provided that $\gamma \ge 1$.


\section{Entanglement entropy for non-Gaussian states}
\label{sec:NGgen}

In this section we extend the calculation of Sec.~\ref{sec:Gauss} to the case of a general non-Gaussian density matrix. As we will show, such a calculation is generally very complicated due to the higher order structure of the underlying correlations. However, under the assumption that non-Gaussian correlations are small, one can make simplifying assumptions which keep the calculation tractable, which are naturally valid in the thin-shell limit \cite{Flynn2023}. Using this approximation, we derive an explicit expression for lowest order correction to the Gaussian reduced density matrix derived in Sec.~\ref{sec:Gauss}, and subsequently use this result to calculate the non-Gaussianity measure introduced in the main text. 

We use the multi-mode Fourier-Weyl relation~\cite{Serafini2017}, which states that the reduced density matrix may be decomposed as,
\begin{align}
\hat{\rho} = \int \frac{d\vb*{\eta}}{\pi^{M_{\mathcal{A}}}} \mathrm{Tr}\left[\hat{D}_{\vb*{\eta}} \hat{\rho}\right]\hat{D}_{-\vb*{\eta}}, \qquad \textrm{where} \qquad \hat{D}_{\vb*{\eta}} = e^{\overline{\sum}_{\vb{k}} \eta_{\vb{k}} \hat{a}^{\dagger}_{\vb{k}}-\eta^*_{\vb{k}} \hat{a}_{\vb{k}}}. 
\end{align} 
Expanding the exponent that defines the Weyl operators $\hat{D}_{\vb*{\eta}}$ using the normal-ordered form, or Glauber-P representation~\cite{Glauber1963,Sudarshan1963, Walls2008}, one finds the equivalent form,
\begin{align}
\hat{\rho} =  \int  \frac{ d\vb*{\eta}}{\pi^{M_{\mathcal{A}}}}  \ \varphi(\vb*{\eta}) e^{\overline{\sum}_{\vb{k}} \eta_{\vb{k}}^* \hat{a}_{\vb{k}} - \overline{\sum}_{\vb{k}} \eta_{\vb{k}} \hat{a}_{\vb{k}}^{\dagger}} e^{-\frac{1}{2} \overline{\sum}_{\vb{k}} \abs{\eta_{\vb{k}}}^2} \qquad \textrm{with} \qquad \varphi(\vb*{\eta}) = \mathrm{Tr}\left[ \hat{\rho} e^{\overline{\sum}_{\vb{k}} \eta_{\vb{k}} \hat{a}_{\vb{k}}^{\dagger}} e^{- \overline{\sum}_{\vb{k}} \eta_{\vb{k}}^* \hat{a}_{\vb{k}}} \right] ,
\label{eq:Prep}
\end{align} 
which will be most convenient for our purposes. Here $\varphi(\vb*{\eta})$ is referred to as the characteristic function. It depends on $M_{\mathcal{A}}$ complex numbers $\eta_{\vb{k}}$, which, following Sec.~\ref{sec:Gauss}, we collect in vectors,
\begin{align}
\vb*{\eta}_{\vb{k}} = \begin{pmatrix} 
    \eta_{\vb{k}}  \\ \eta_{-\vb{k}} \\  \eta_{\vb{k}}^*  \\ \eta_{-\vb{k}}^*
\end{pmatrix}.
\end{align}
The $2 M_{\mathcal{A}}$ dimensional vector $\vb*{\eta}$ then collects the vectors $\vb*{\eta}_{\vb{k}}$ for all $\frac{1}{2} M_{\mathcal{A}}$ pairs $(\vb{k}, -\vb{k})$. We have also introduced the multidimensional integral,
\begin{align}
\begin{split}
\int \frac{d \vb*{\eta}}{\pi^{M_{\mathcal{A}}}} \equiv \overline{\prod_{\vb{k}>0}} \int \frac{d\vb*{\eta}_{\vb{k}}}{\pi^2} \equiv \overline{\prod_{\vb{k}>0}} \frac{1}{\pi^2} \int_{-\infty}^{\infty} d\mathrm{Re}(\eta_{\vb{k}}) \int_{-\infty}^{\infty} d\mathrm{Im}(\eta_{\vb{k}}) \int_{-\infty}^{\infty} d\mathrm{Re}(\eta_{-\vb{k}}) \int_{-\infty}^{\infty} d\mathrm{Im}(\eta_{-\vb{k}}).
\end{split}
\end{align}

\subsection{Cumulant expansion of characteristic function}

For our purposes, the characteristic function $\varphi(\vb*{\eta})$ as introduced in Eq.~\eqref{eq:Prep} is most conveniently expressed as the expectation value,
\begin{align}
\varphi(\vb*{\eta}) = \expval*{e^{\overline{\sum}_{\vb{k}} \eta_{\vb{k}} \hat{a}_{\vb{k}}^{\dagger}} e^{- \overline{\sum}_{\vb{k}} \eta_{\vb{k}}^* \hat{a}_{\vb{k}}}}.
\end{align}
as it can be directly related to the cumulant expansion of expectation values in subspace $\mathcal{A}$ \cite{Kira2008}. To see this, consider the correlation-generating function $\Xi(\vb*{\eta}) = \ln \varphi(\vb*{\eta})$, which is analytic near $\vb*{\eta} = \vb*{0}$, with $\Xi(\vb*{0}) = 0$ specifically. Knowing this, we can write out the Taylor expansion of  $\Xi(\vb*{\eta})$ around $\vb*{\eta} = \vb*{0}$, after which we use Eq.~\eqref{eq:CumDef} to obtain,
\begin{align}
\begin{split}
\Xi(\vb*{\eta}) =  \sum_{\bar{\vb{n}}} \sum_{\vb{n}} \expval*{\prod_{i = 1}^{M_{\mathcal{A}}} (\hat{a}_{\vb{k}_i}^{\dagger})^{\bar{n}_i} \prod_{j = 1}^{M_{\mathcal{A}}} (\hat{a}_{\vb{q}_j})^{n_j}}_{\mathrm{c}} \prod_{i = 1}^{M_{\mathcal{A}}} \frac{\left(\eta_{\vb{k}_i}\right)^{\bar{n}_i}}{\bar{n}_i !} \prod_{j = 1}^{M_{\mathcal{A}}} \frac{\left(-\eta_{\vb{k}_j}^*\right)^{n_j}}{n_j !},
\end{split}
\label{eq:WignerFuncExp}
\end{align}
where now $\bar{\vb{n}}$ and $\vb{n}$ are $M_{\mathcal{A}}$ dimensional vectors of non-negative integers, and all unique such vectors are summed over to produce the Taylor series. Hence, we observe that the characteristic function can be written in terms of a cumulant series, where higher order terms correspond with higher powers of the coefficients $\eta_{\vb{k}}$. In terms of the three correlations in space $\mathcal{A}$ included in our model,
\begin{align}
\begin{split}
\Xi(\vb*{\eta}) = - \overline{\sum_{\vb{k}}} n_{\vb{k}} \abs{\eta_{\vb{k}}}^2 + \frac{1}{2}\overline{\sum_{\vb{k}}} \kappa_{\vb{k}}^* \eta_{\vb{k}} \eta_{-\vb{k}} + \frac{1}{2} \overline{\sum_{\vb{k}}} \kappa_{\vb{k}} \eta_{\vb{k}}^* \eta_{-\vb{k}}^* + \frac{1}{6}\overline{\sum_{\vb{k}, \vb{q}}}  \eta_{\vb{k}} \eta_{\vb{q}} \eta_{-\vb{k} - \vb{q}} R_{\vb{k}, \vb{q}}^*  -\frac{1}{6}\overline{\sum_{\vb{k}, \vb{q}}}  \eta_{\vb{k}}^* \eta_{\vb{q}}^* \eta_{-\vb{k} - \vb{q}}^* R_{\vb{k}, \vb{q}}.
\end{split}
\label{eq:WignerFuncExp2}
\end{align}
Here special care has to be taken in computing the prefactors. For $\kappa_{\vb{k}}$, the sum is multiplied by a factor $\frac{1}{2}$ to account for the fact that the number of pairs $(\vb{k}, -\vb{k})$ in the shell is half the total number of modes $M_{\mathcal{A}}$. For the triplet the situation is slightly more complicated. For example, straightforward application of Eq.~\eqref{eq:WignerFuncExp} gives the contribution $\Xi_{R^*}(\vb*{\eta})$ to $\Xi(\vb*{\eta})$ due to $R_{\vb{k} , \vb{q}}^*$ as,
\begin{align}
\begin{split}
\Xi_{R^*}(\vb*{\eta}) =   \sum_{\vb{k} \in Q_1} \sum_{\vb{q} \in Q_2} \eta_{\vb{k}} \eta_{\vb{q}} \eta_{-\vb{k} - \vb{q}} R_{\vb{k}, \vb{q}}^* + \frac{1}{2} \overline{\sum_{\vb{k}}} \eta_{\vb{k}}^2 \eta_{-2\vb{k}} R_{\vb{k}, \vb{q}}^*,
\end{split}
\end{align}
where $Q_i$ denotes the $i$'th quadrant of the shell $\mathcal{A}$. The second term accounts for the fact that inside the triplet correlation two atoms may be in the same momentum state, multiplied by a factor $1/2$ from the Taylor expansion. As there are 6 choices for the different configurations of quadrants, we can also write,
\begin{align}
\begin{split}
\Xi_{R^*}(\vb*{\eta}) &=  \frac{1}{6}\overline{\sum_{\vb{k}, \vb{q}}} \eta_{\vb{k}} \eta_{\vb{q}} \eta_{-\vb{k} - \vb{q}} R_{\vb{k}, \vb{q}}^* \bar{\delta}_{\vb{k}, \vb{q}}  + \frac{1}{2} \overline{\sum_{\vb{k}}} \eta_{\vb{k}}^2 \eta_{-2\vb{k}} R_{\vb{k}, \vb{q}}^*,
\end{split}
\end{align}
where $\bar{\delta}_{\vb{k}, \vb{q}} = (1 - \delta_{\vb{k}, \vb{q}}) (1 - \delta_{\vb{k}, -\vb{k} - \vb{q}}) (1 - \delta_{\vb{q}, -\vb{k} - \vb{q}})$, and we have used the inherent permutation symmetries of the $R_{\vb{k}, \vb{q}}$ cumulant. Since the second sum exactly accounts for the ``missing" contributions to the first sum due to the delta functions, we can simply take the two sums together, after which we obtain the contribution in Eq.~\eqref{eq:WignerFuncExp2}.

\begin{figure}
\includegraphics[width=0.6\linewidth]{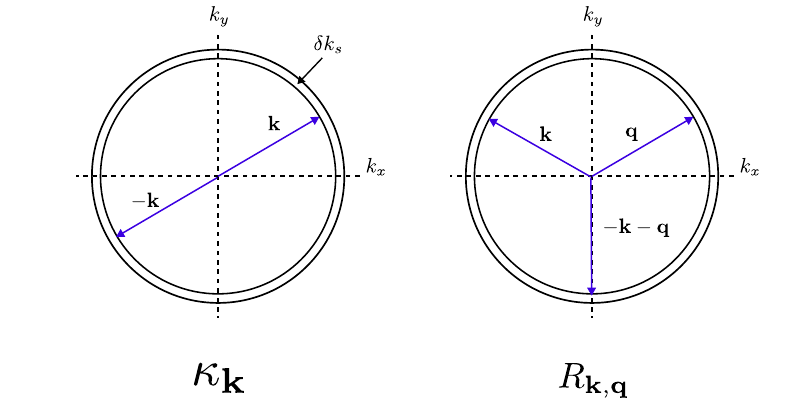}
\caption{\label{fig:shellconfigs} Allowed vector configurations of $\kappa_{\vb{k}}$ and $R_{\vb{k}, \vb{q}}$ in the thin-shell limit, projected on the $k_z = 0$ plane. In $\kappa_{\vb{k}}$, two momentum vectors with length $\abs{\vb*{k}} = k_s$ point to opposite ends of the shell. In $R_{\vb{k}, \vb{q}}$, three momentum vectors with length $\abs*{\vb{k}} = \abs*{\vb{q}} = \abs*{\vb{k}+\vb{q}} = k_s$ point to the shell, with relative angles of $\theta_s = 2 \pi/3$.}
\end{figure}

To obtain the reduced density matrix, we should now insert Eq.~\eqref{eq:WignerFuncExp2} back into Eq.~\eqref{eq:Prep}, and solve the resulting integral. Such a calculation however, is very complicated due to the cubic nature of $R_{\vb{k}, \vb{q}}$. To simplify the problem to a tractable form, we assume that the non-Gaussian correlations are much smaller than the Gaussian correlations, after which we may adopt the following linearization,
\begin{align}
\begin{split}
\varphi(\vb*{\eta}) &\approx e^{\Xi_G(\vb*{\eta})} \left[1 + \frac{1}{6}\overline{\sum_{\vb{k}, \vb{q}}}  \eta_{\vb{k}} \eta_{\vb{q}} \eta_{-\vb{k} - \vb{q}} R_{\vb{k}, \vb{q}}^*  -\frac{1}{6}\overline{\sum_{\vb{k}, \vb{q}}}  \eta_{\vb{k}}^* \eta_{\vb{q}}^* \eta_{-\vb{k} - \vb{q}}^* R_{\vb{k}, \vb{q}}  \right].
\end{split}
\label{eq:CharFuncApprox}
\end{align}
where $\Xi_G(\vb*{\eta})$ is the Gaussian part of $\Xi(\vb*{\eta})$, corresponding to the first three terms in Eq.~\eqref{eq:WignerFuncExp2} and dependent on $n_{\vb{k}}$ and $\kappa_{\vb{k}}$. A similar simplification of the characteristic function was recently shown to be accurate in modeling the reduced density matrix of interacting fermions \cite{Froland2024}, again assuming the non-Gaussian correlations are small. Here, we point out that in the thin-shell approximation this assumption is automatically fulfilled, which is seen directly by comparing the sum over doublet and triplet correlations, and taking the thermodynamic limit (see Sec.~\ref{sec:model}),
\begin{align}
\overline{\sum_{\vb{k}}} \kappa_{\vb{k}} \rightarrow \frac{V}{2 \pi^2} k_s^2 \delta k_s \ \kappa(k_s), \qquad
\overline{\sum_{\vb{k}, \vb{q}}} R_{\vb{k}, \vb{q}} \rightarrow \frac{V^2}{8 \pi^4} k_s^4 (\delta k_s)^2 \ R(k_s, k_s, \theta_s),
\label{eq:ThermoLim}
\end{align}
 with $\theta_s = 2 \pi/3$ the angle between the vectors in the triplet, see also Fig.~\ref{fig:shellconfigs}. Evidently, the magnitude of the non-Gaussian correlation scales quadratically with the shell width $\delta k_s$, while the Gaussian correlations scale linearly, justifying an expansion in this small parameter. In the lowest order approximation, we simply omit the non-Gaussian contributions to the characteristic function, after which one obtains the thin-shell approximation as used in the main text. Note that this argument also extends to beyond third order cumulants, whose integrals scale with successively higher powers of the shell width. Hence, the thin shell partition provides a generally efficient way to approximate the characteristic function, also outside quench scenarios, as was already shown for interacting Fermi gases in Ref.~\cite{Flynn2023}. 

\subsection{Beyond-Gaussian correction to the reduced density matrix}

We now derive the non-Gaussian correction $\hat{\rho}'$ to the Gaussian reduced density matrix $\hat{\rho}_G$, for which we will use the characteristic function in Eq.~\eqref{eq:CharFuncApprox}. Upon inserting into Eq.~\eqref{eq:Prep}. we obtain,
\begin{align}
\begin{split}
\hat{\rho} &=  \int  \frac{ d\vb*{\eta}}{\pi^{M_{\mathcal{A}}}}  \ e^{\Xi_G(\vb*{\eta})} e^{\overline{\sum}_{\vb{k}} \eta_{\vb{k}}^* \hat{a}_{\vb{k}} - \overline{\sum}_{\vb{k}} \eta_{\vb{k}} \hat{a}_{\vb{k}}^{\dagger}} e^{-\frac{1}{2} \overline{\sum}_{\vb{k}} \abs{\eta_{\vb{k}}}^2} \\ &+ \frac{1}{6} \overline{\sum_{\vb{k}, \vb{q}}} R_{\vb{k}, \vb{q}}^*  \int  \frac{d\vb*{\eta}}{\pi^{M_{\mathcal{A}}}}  \ \eta_{\vb{k}} \eta_{\vb{q}} \eta_{-\vb{k} - \vb{q}} e^{\Xi_G(\vb*{\eta})} e^{\overline{\sum}_{\vb{k}} \eta_{\vb{k}}^* \hat{a}_{\vb{k}} - \overline{\sum}_{\vb{k}} \eta_{\vb{k}} \hat{a}_{\vb{k}}^{\dagger}} e^{-\frac{1}{2} \overline{\sum}_{\vb{k}} \abs{\eta_{\vb{k}}}^2} \\ &- \frac{1}{6} \overline{\sum_{\vb{k}, \vb{q}}} R_{\vb{k}, \vb{q}}  \int  \frac{ d\vb*{\eta}}{\pi^{M_{\mathcal{A}}}}  \ \eta_{\vb{k}}^* \eta_{\vb{q}}^* \eta_{-\vb{k} - \vb{q}}^* e^{\Xi_G(\vb*{\eta})} e^{\overline{\sum}_{\vb{k}} \eta_{\vb{k}}^* \hat{a}_{\vb{k}} - \overline{\sum}_{\vb{k}} \eta_{\vb{k}} \hat{a}_{\vb{k}}^{\dagger}} e^{-\frac{1}{2} \overline{\sum}_{\vb{k}} \abs{\eta_{\vb{k}}}^2},
\end{split}
\end{align}
which motivates the identification $\hat{\rho} = \hat{\rho}_G + \hat{\rho}'$ in the main text, where $\hat{\rho}_G$ corresponds with the first line, and $\hat{\rho}'$ with the second and third lines. To evaluate the expression above, we recognize that $\Xi_G(\vb*{\eta}) = \overline{\sum}_{\vb{k} > 0} \xi_G(\vb*{\eta}_{\vb{k}})$, where,
\begin{align}
\begin{split}
\xi_G(\vb*{\eta}_{\vb{k}}) = -\frac{1}{2} \vb*{\eta}_{\vb{k}}^{\dagger}  \Omega C_{\vb{k}} \Omega \vb*{\eta}_{\vb{k}} + \frac{1}{4} \vb*{\eta}_{\vb{k}}^{\dagger} \vb*{\eta}_{\vb{k}},
\end{split}
\end{align}
with $C_{\vb{k}}$ as defined in Eq.~\eqref{eq:Cdef}. Hence the integrals can be factorized as,
\begin{align}
\begin{split}
\hat{\rho}_G &= \overline{\prod_{\vb{k} > 0}}  \int  \frac{ d\vb*{\eta}_{\vb{k}}}{\pi^{2}}  \ e^{-\frac{1}{2} \vb*{\eta}_{\vb{k}}^{\dagger} \Omega C_{\vb{k}} \Omega \vb*{\eta}_{\vb{k}}} e^{- \hat{\vb{\alpha}}_{\vb{k}}^{\dagger} \Omega \vb*{\eta}_{\vb{k}}} ,
\end{split}
\end{align}
and, writing $\rho' = \hat{\sigma} + \hat{\sigma}^{\dagger}$,
\begin{align}
\begin{split}
\hat{\sigma} &= \frac{1}{6} \overline{\sum_{\vb{k}, \vb{q}}} R_{\vb{k}, \vb{q}}^* \bar{\delta}_{\vb{k}, \vb{q}} \  \underset{\substack{\vb{p} \ne \pm \vb{k} , \pm\vb{q}, \pm(\vb{k} + \vb{q}) \\ \vb{p > 0} }}{\overline{\prod}} \int  \frac{d\vb*{\eta}_{\vb{p}}}{\pi^{2}}  \  e^{-\frac{1}{2} \vb*{\eta}_{\vb{p}}^{\dagger} \Omega C_{\vb{p}} \Omega  \vb*{\eta}_{\vb{p}}} e^{- \hat{\vb{\alpha}}_{\vb{p}}^{\dagger} \Omega \vb*{\eta}_{\vb{p}}}  \\ & \qquad \qquad \qquad \times \prod_{\vb{l} = \vb{k}, \vb{q}, -\vb{k} - \vb{q}}  \int  \frac{d\vb*{\eta}_{\vb{l}}}{\pi^{2}}  \ \eta_{\vb{l}}  e^{-\frac{1}{2} \vb*{\eta}_{\vb{l}}^{\dagger}  \Omega C_{\vb{l}} \Omega \vb*{\eta}_{\vb{l}}} e^{- \hat{\vb{\alpha}}_{\vb{l}}^{\dagger} \Omega \vb*{\eta}_{\vb{l}}} \\ & + \frac{1}{2} \overline{\sum_{\vb{k}}} R_{\vb{k}, \vb{k}}^* \underset{\substack{\vb{p} \ne \pm \vb{k}, \pm 2\vb{k} \\ \vb{p} > 0}}{\overline{\prod}}   \int  \frac{d\vb*{\eta}_{\vb{p}}}{\pi^{2}}  \  e^{-\frac{1}{2} \vb*{\eta}_{\vb{p}}^{\dagger}  \Omega C_{\vb{p}} \Omega \vb*{\eta}_{\vb{p}}} e^{- \hat{\vb{\alpha}}_{\vb{p}}^{\dagger} \Omega \vb*{\eta}_{\vb{p}}}  \\ & \qquad \qquad \qquad \times \int \frac{d \eta_{\vb{k}}}{\pi^2} \eta_{\vb{k}}^2  e^{- \frac{1}{2} \vb*{\eta}_{\vb{k}}^{\dagger} \Omega C_{\vb{k}} \Omega  \vb*{\eta}_{\vb{k}}} e^{- \hat{\vb{\alpha}}_{\vb{k}}^{\dagger} \Omega \vb*{\eta}_{\vb{k}}}  \int \frac{d \eta_{-2\vb{k}}}{\pi^2} \eta_{-2\vb{k}}  e^{-\frac{1}{2} \vb*{\eta}_{-2\vb{k}}^{\dagger} \Omega C_{-2\vb{k}} \Omega  \vb*{\eta}_{-2\vb{k}}} e^{- \hat{\vb{\alpha}}_{-2\vb{k}}^{\dagger} \Omega \vb*{\eta}_{-2\vb{k}}}. 
\end{split}
\label{eq:rhoPGINT}
\end{align}
Here we have split the product over momentum states to separate the integrals that are identical to those found in the Gaussian state, i.e. those containing the momentum index $\vb{p}$, and the new integrals containing powers of the coherent state coefficients. We have also split the momentum sum in front to separate those terms in which two momenta are equal, since they lead to a different Gaussian integral. Note that $R_{\vb{k}, -\vb{k}} = 0$ by construction.

In all the integrals above we now use the transformation $U_{\vb{k}}$ defined in Eq.~\eqref{eq:Uk} to change the variable of integration as $\vb*{\eta}_{\vb{k}} = U_{\vb{k}} \vb*{\gamma}_{\vb{k}}$, which preserves the integration volume by its symplectic property $\abs{u_{\vb{k}}}^2 - \abs{v_{\vb{k}}}^2 = 1$. Then the Gaussian part of the density matrix becomes,
\begin{align}
\begin{split}
\hat{\rho}_G &= \overline{\prod_{\vb{k} > 0}}  \int  \frac{ d\vb*{\gamma}_{\vb{k}}}{\pi^{2}}  \ e^{-\frac{1}{2} \vb*{\gamma}_{\vb{k}}^{\dagger} U_{\vb{k}}^{\dagger} \Omega C_{\vb{k}} \Omega U_{\vb{k}} \vb*{\gamma}_{\vb{k}}} e^{- \hat{\vb*{\beta}}_{\vb{k}}^{\dagger} U_{\vb{k}}^{\dagger} \Omega U_{\vb{k}} \vb*{\gamma}_{\vb{k}}} \\
&= \overline{\prod_{\vb{k} > 0}}  \int  \frac{ d\vb*{\gamma}_{\vb{k}}}{\pi^{2}}  \ e^{-\frac{1}{2} \vb*{\gamma}_{\vb{k}}^{\dagger} \tilde{N}_{\vb{k}} \vb*{\gamma}_{\vb{k}} - \frac{1}{4} \vb*{\gamma}_{\vb{k}}^{\dagger} \vb*{\gamma}_{\vb{k}}} e^{- \hat{\vb*{\beta}}_{\vb{k}}^{\dagger} \Omega \vb*{\gamma}_{\vb{k}}} 
\end{split}
\label{eq:rhoGbtrans}
\end{align}
where in going to the second line we have used the relations derived in Sec.~\ref{sec:Gauss}. We have defined $\tilde{N}_{\vb{k}}$ as a $4 \times 4$ diagonal matrix containing the associated mode occupation number $\tilde{n}_{\vb{k}}$ along the diagonal. 

To transform back to an operatorial representation, we  introduce a set of $\pm \vb{k}$ bimodal coherent states in the entanglement eigenmode basis, defined as \cite{Serafini2017},
\begin{align}
\ket*{\vb*{\eta}_{\vb{k}}} = e^{\vb*{\beta}_{\vb{k}}^{\dagger} \Omega \vb*{\eta}_{\vb{k}}} \ket*{0}_{\vb{k}}, \qquad \int \frac{d \vb*{\eta}_{\vb{k}}}{\pi^2} \dyad*{\vb*{\eta}_{\vb{k}}} = \mathbb{I}_{\vb{k}} \mathbb{I}_{-\vb{k}},
\end{align}
where $\ket*{0}_{\vb{k}}$ is the vacuum for the $\pm \vb{k}$ entanglement eigenmodes, and $\mathbb{I}_{\vb{k}}$ the identity operator in the Fock space for eigenmode $\vb{k}$. Note here that each mode is independent, i.e. $\dyad*{\vb*{\eta}_{\vb{k}}}$ acts solely in the $\pm \vb{k}$ Fock space. Upon inserting a complete set of coherent states into Eq.~\eqref{eq:rhoGbtrans}, and using the fact that $\ket*{\vb*{\eta}_{\vb{k}}}$ is an eigenstate of both $b_{\vb{k}}$ and $b_{-\vb{k}}$, we obtain,
\begin{align}
\begin{split}
\hat{\rho}_G &=  \overline{\prod_{\vb{k} > 0}}  \int  \frac{ d\vb*{\gamma}_{\vb{k}} d\vb*{\eta}_{\vb{k}}}{\pi^{4}}  \ e^{-\frac{1}{2} \vb*{\gamma}_{\vb{k}}^{\dagger} \tilde{N}_{\vb{k}} \vb*{\gamma}_{\vb{k}}} e^{- \hat{\vb*{\eta}}_{\vb{k}}^{\dagger} \Omega \vb*{\gamma}_{\vb{k}}} \dyad*{\vb*{\eta}_{\vb{k}}} \\
&= \overline{\prod_{\vb{k}>0}} \frac{1}{\tilde{n}_{\vb{k}}^2}  \int  \frac{ d\vb*{\eta}_{\vb{k}}}{\pi^{2}}  \ e^{- \frac{1}{2\tilde{n}_{\vb{k}}} \vb*{\eta}_{\vb{k}}^{\dagger} \vb*{\eta}_{\vb{k}} } \dyad*{\vb*{\eta}_{\vb{k}}} \\
&= \frac{1}{\overline{\prod}_{\vb{k}} (1 + \tilde{n}_{\vb{k}})} e^{- \overline{\sum}_{\vb{k}} \lambda_{\vb{k}} \hat{b}_{\vb{k}}^{\dagger} \hat{b}_{\vb{k}}},
\end{split}
\end{align}
where in going from the first to the second line we have solved a Gaussian integral, and in going from the second to the third line we have used the identity,
\begin{align}
\begin{split}
e^{- \frac{1}{2\tilde{n}_{\vb{k}}} \vb*{\eta}_{\vb{k}}^{\dagger} \vb*{\eta}_{\vb{k}} } \dyad*{\vb*{\eta}_{\vb{k}}} = e^{-\frac{1}{2} \lambda_{\vb{k}} \left(\hat{b}_{\vb{k}}^{\dagger} \hat{b}_{\vb{k}} + \hat{b}_{-\vb{k}}^{\dagger} \hat{b}_{-\vb{k}} \right)} \dyad*{e^{\frac{1}{2} \lambda_{\vb{k}}}\vb*{\eta}_{\vb{k}}} e^{-\frac{1}{2} \lambda_{\vb{k}} \left(\hat{b}_{\vb{k}}^{\dagger} \hat{b}_{\vb{k}} + \hat{b}_{-\vb{k}}^{\dagger} \hat{b}_{-\vb{k}} \right)}.
\end{split}
\end{align}
So as expected, we retrieve the result already derived in Sec.~\ref{sec:Gauss}. We now repeat the procedure for the non-Gaussian correction first transforming the integration variable using $U_{\vb{k}}$, inserting a complete set of coherent states in the entanglement eigenmode basis, and solving the resulting Gaussian integral. Starting from Eq.~\eqref{eq:rhoPGINT}, we find,
\begin{align}
\begin{split}
\hat{\sigma} &= \frac{1}{6} \overline{\sum_{\vb{k}, \vb{q}}} R_{\vb{k}, \vb{q}}^* \bar{\delta}_{\vb{k}, \vb{q}} \  \underset{\substack{\vb{p} \ne \pm \vb{k} , \pm\vb{q}, \pm(\vb{k} + \vb{q}) \\ \vb{p > 0} }}{\overline{\prod}} \frac{1}{\tilde{n}_{\vb{p}}}\int  \frac{ d\vb*{\eta}_{\vb{p}}}{\pi^{2}}  \ e^{- \frac{1}{2\tilde{n}_{\vb{k}}} \vb*{\eta}_{\vb{p}}^{\dagger} \vb*{\eta}_{\vb{p}} } \dyad*{\vb*{\eta}_{\vb{p}}}  \\ & \qquad \qquad \qquad \times \prod_{\vb{l} = \vb{k}, \vb{q}, -\vb{k} - \vb{q}}  \frac{1}{\tilde{n}_{\vb{l}}^3} \int  \frac{d\vb*{\eta}_{\vb{l}}}{\pi^{2}}  \  (u_{\vb{l}} \eta_{\vb{l}} - v_{\vb{l}}^* \eta_{-\vb{l}}^*)  e^{- \frac{1}{2\tilde{n}_{\vb{l}}} \vb*{\eta}_{\vb{l}}^{\dagger} \vb*{\eta}_{\vb{l}} } \dyad*{\vb*{\eta}_{\vb{l}}} \\ & + \frac{1}{2} \overline{\sum_{\vb{k}}} R_{\vb{k}, \vb{k}}^* \underset{\substack{\vb{p} \ne \pm \vb{k}, \pm 2\vb{k} \\ \vb{p} > 0}}{\overline{\prod}} \frac{1}{\tilde{n}_{\vb{p}}}  \int  \frac{d\vb*{\eta}_{\vb{p}}}{\pi^{2}}  \  e^{- \frac{1}{2\tilde{n}_{\vb{p}}} \vb*{\eta}_{\vb{p}}^{\dagger} \vb*{\eta}_{\vb{p}} } \dyad*{\vb*{\eta}_{\vb{p}}}  \\ & \qquad \qquad \qquad \times \frac{1}{\tilde{n}_{\vb{k}}^4} \int \frac{d \vb*{\eta}_{\vb{k}}}{\pi^2} (u_{\vb{k}}^2 \eta_{\vb{k}}^2 + (v_{\vb{k}}^*)^2 (\eta_{-\vb{k}}^*)^2 - 2 u_{\vb{k}} v_{\vb{k}}^* \eta_{\vb{k}} \eta_{-\vb{k}}^*) e^{- \frac{1}{2\tilde{n}_{\vb{k}}} \vb*{\eta}_{\vb{k}}^{\dagger} \vb*{\eta}_{\vb{k}} } \dyad*{\vb*{\eta}_{\vb{k}}} \\ & \qquad \qquad \qquad \times \frac{1}{\tilde{n}_{2\vb{k}}^3} \int \frac{d \vb*{\eta}_{-2\vb{k}}}{\pi^2} (u_{2\vb{k}} \eta_{-2\vb{k}} - v_{2\vb{k}}^* \eta_{2\vb{k}}^*) e^{- \frac{1}{2\tilde{n}_{2\vb{k}}} \vb*{\eta}_{-2\vb{k}}^{\dagger} \vb*{\eta}_{-2\vb{k}} } \dyad*{\vb*{\eta}_{-2\vb{k}}}.
\end{split}
\label{eq:rhopPhase}
\end{align}
Again we can write this as a function of eigenmode operators, noting that,
\begin{align}
\begin{split}
&\int  \frac{d\vb*{\eta}_{\vb{l}}}{\pi^{2}}  \  (u_{\vb{l}} \eta_{\vb{l}} - v_{\vb{l}}^* \eta_{-\vb{l}}^*)  e^{- \frac{1}{2\tilde{n}_{\vb{l}}} \vb*{\eta}_{\vb{l}}^{\dagger} \vb*{\eta}_{\vb{l}} } \dyad*{\vb*{\eta}_{\vb{l}}} = e^{-\frac{5}{2} \lambda_{\vb{l}}} e^{- \lambda_{\vb{l}} (\hat{b}_{\vb{l}}^{\dagger} \hat{b}_{\vb{l}} +  \hat{b}_{-\vb{l}}^{\dagger} \hat{b}_{-\vb{l}})} \left(e^{-\frac{1}{2} \lambda_{\vb{l}}} u_{\vb{l}} \hat{b}_{\vb{l}} - e^{\frac{1}{2}\lambda_{\vb{l}}} v_{\vb{l}}^* \hat{b}_{-\vb{l}}^{\dagger} \right) \\
&\int \frac{d \vb*{\eta}_{\vb{k}}}{\pi^2} (u_{\vb{k}}^2 \eta_{\vb{k}}^2 + (v_{\vb{k}}^*)^2 (\eta_{-\vb{k}}^*)^2 - 2 u_{\vb{k}} v_{\vb{k}}^* \eta_{\vb{k}} \eta_{-\vb{k}}^*) e^{- \frac{1}{2\tilde{n}_{\vb{k}}} \vb*{\eta}_{\vb{k}}^{\dagger} \vb*{\eta}_{\vb{k}} } \dyad*{\vb*{\eta}_{\vb{k}}} = e^{-3 \lambda_{\vb{k}}} e^{- \lambda_{\vb{l}} (\hat{b}_{\vb{l}}^{\dagger} \hat{b}_{\vb{l}} +  \hat{b}_{-\vb{l}}^{\dagger} \hat{b}_{-\vb{l}})}  \left(e^{-\frac{1}{2}\lambda_{\vb{l}}} u_{\vb{l}} \hat{b}_{\vb{l}} - e^{\frac{1}{2}\lambda_{\vb{l}}} v_{\vb{l}}^* \hat{b}_{-\vb{l}}^{\dagger} \right)^2.
\end{split}
\end{align}
Inserting back into $\hat{\rho}'$, we then recognize that the second sum again compensates the ``missing" terms in the first sum, such that we can take them together to finally obtain,
\begin{align}
\begin{split}
\hat{\sigma} = \hat{\rho}_G \times \frac{1}{6} \overline{\sum_{\vb{k}, \vb{q}}} R_{\vb{k}, \vb{q}} \prod_{\vb{l} = \vb{k}, \vb{q}, -\vb{k} - \vb{q}} \frac{1}{\sqrt{\tilde{n}_{\vb{l}} (1 + \tilde{n}_{\vb{l}})}} \left(e^{-\frac{1}{2}\lambda_{\vb{l}}} u_{\vb{l}} \hat{b}_{\vb{l}} - e^{\frac{1}{2}\lambda_{\vb{l}}} v_{\vb{l}}^* \hat{b}_{-\vb{l}}^{\dagger} \right)
\end{split}
\label{eq:rhopOp}
\end{align}
Evidently this operator vanishes on the diagonal in the eigenmode Fock space, implying that $\mathrm{Tr} [\hat{\rho}'] = 0$. Beyond this simple observation however, the operatorial form in Eq.~\eqref{eq:rhopOp} is generally much more difficult to work with than the coherent state representation in Eq.~\eqref{eq:rhopPhase}, which we will use going forward.

\subsection{Non-Gaussianity measure}
\label{sec:NGmeas}

As discussed in the main text, we quantify the validity of the thin-shell approximation through the non-Gaussianity measure $\mathcal{N}(\hat{\rho} || \hat{\rho}_G)$ \cite{Genoni2007, Genoni2008, Baek2018, Park2021}, expressed in terms of traces over products of density matrices, which can be directly evaluated in the coherent state representation derived in the previous section. First, we note that $\hat{\rho}'$ is a purely off-diagonal operator in the Fock space defined by the eigenmodes $\hat{b}_{\vb{k}}$, which means that $\mathrm{Tr} [\hat{\rho}'] = \mathrm{Tr} [\hat{\rho}_G \hat{\rho}'] = 0$. Hence we can rewrite the non-Gaussianity as,
\begin{align}
\begin{split}
\mathcal{N}(\hat{\rho} || \hat{\rho}_G)= \frac{1}{2} \frac{\mathrm{Tr}[(\hat{\rho}')^2]}{ \mathrm{Tr} [\hat{\rho}_G^2] +  \mathrm{Tr} [(\hat{\rho}')^2]}.
\end{split} 
\end{align}
As an illustrative example, we will evaluate $\mathrm{Tr} [\hat{\rho}_G^2]$ in the coherent state representation,
\begin{align}
\begin{split}
\mathrm{Tr} [\hat{\rho}_G^2] &= \mathrm{Tr} \left[ \overline{\prod_{\vb{k} > 0}} \int \frac{d \vb*{\eta}_{\vb{k}}}{ \pi^2} \int \frac{d \vb*{\phi}_{\vb{k}}}{ \pi^2}  e^{- \frac{1}{2\tilde{n}_{\vb{k}}} \vb*{\eta}_{\vb{k}}^{\dagger} \vb*{\eta}_{\vb{k}} } e^{- \frac{1}{2\tilde{n}_{\vb{k}}} \vb*{\phi}_{\vb{k}}^{\dagger} \vb*{\phi}_{\vb{k}}} \ket*{\eta_{\vb{k}}} \braket{\eta_{\vb{k}}}{\phi_{\vb{k}}} \bra*{\phi_{\vb{k}}} \right] \\
&=  \overline{\prod_{\vb{k} > 0}} \int \frac{d \vb*{\gamma}_{\vb{k}}}{ \pi^2}  \int \frac{d \vb*{\eta}_{\vb{k}}}{ \pi^2} \int \frac{d \vb*{\phi}_{\vb{k}}}{ \pi^2} e^{-\frac{1}{\tilde{n}_{\vb{k}}} \left(\abs{\eta_{\vb{k}}}^2 + \abs{\eta_{-\vb{k}}}^2 \right)} e^{-\frac{1}{\tilde{n}_{\vb{k}}} \left(\abs{\phi_{\vb{k}}}^2 + \abs{\phi_{-\vb{k}}}^2 \right)} \braket*{\gamma_{\vb{k}}}{\eta_{\vb{k}}} \braket{\eta_{\vb{k}}}{\phi_{\vb{k}}} \braket*{\phi_{\vb{k}}}{\gamma_{\vb{k}}}.
\end{split} 
\label{eq:rhoG2deriv1}
\end{align}
Here we have used the fact that the trace factorizes over all momentum modes. Now, using the coherent state overlap,
\begin{align}
\begin{split}
\braket*{\vb*{\gamma}_{\vb{k}}}{\vb*{\eta}_{\vb{k}}} = e^{-\frac{1}{2} (\abs{\gamma_{\vb{k}}}^2 + \abs{\eta_{\vb{k}}}^2   - 2 \gamma_{\vb{k}}^* \eta_{\vb{k}} )} e^{-\frac{1}{2} (\abs{\gamma_{-\vb{k}}}^2 + \abs{\eta_{-\vb{k}}}^2   - 2 \gamma_{-\vb{k}}^* \eta_{-\vb{k}} )}, 
\end{split}
\label{eq:overlap}
\end{align}
we can calculate the integral over $\gamma_{\vb{k}}$,
\begin{align}
\begin{split}
&\int \frac{d \vb*{\gamma}_{\vb{k}}}{\pi^2}  \braket*{\vb*{\gamma}_{\vb{k}}}{\vb*{\eta}_{\vb{k}}}  \braket*{\vb*{\eta}_{\vb{k}}}{\vb*{\phi}_{\vb{k}}} \braket*{\vb*{\phi}_{\vb{k}}}{\vb*{\gamma}_{\vb{k}}} = e^{-\abs*{\eta_{\vb{k}}}^2 + \eta_{\vb{k}}^* \phi_{\vb{k}} + \eta_{\vb{k}} \phi_{\vb{k}}^*} e^{-\abs*{\eta_{-\vb{k}}}^2 + \eta_{-\vb{k}}^* \phi_{-\vb{k}} + \eta_{-\vb{k}} \phi_{-\vb{k}}^*}  e^{-\abs*{\phi_{\vb{k}}}^2} e^{-\abs*{\phi_{-\vb{k}}}^2} .
\end{split}
\end{align}
Substituting back into Eq.~\eqref{eq:rhoG2deriv1} one obtains,
\begin{align}
\begin{split}
\mathrm{Tr} [\hat{\rho}_G^2] &=  \overline{\prod_{\vb{k} > 0}} \frac{1}{\tilde{n}_{\vb{k}}^4}  \int \frac{d \vb*{\eta}_{\vb{k}}}{ \pi^2} \int \frac{d \vb*{\phi}_{\vb{k}}}{ \pi^2} 
e^{-\left(\frac{1}{\tilde{n}_{\vb{k}}}+1\right)\abs{\eta_{\vb{k}}}^2 + \eta_{\vb{k}}^* \phi_{\vb{k}} + \eta_{\vb{k}} \phi_{\vb{k}}^*} e^{-\left(\frac{1}{\tilde{n}_{\vb{k}}}+1\right)\abs{\eta_{-\vb{k}}}^2 + \eta_{-\vb{k}}^* \phi_{-\vb{k}} + \eta_{-\vb{k}} \phi_{-\vb{k}}^*}
\\ & \qquad \qquad \qquad \qquad \times e^{-\left(\frac{1}{\tilde{n}_{\vb{k}}}+1\right)\abs{\phi_{\vb{k}}}^2} e^{-\left(\frac{1}{\tilde{n}_{\vb{k}}}+1\right)\abs{\phi_{-\vb{k}}}^2} \\
&= \overline{\prod_{\vb{k} > 0}} \frac{1}{\tilde{n}_{\vb{k}}^4}  \left(\frac{\tilde{n}_{\vb{k}}}{1 + \tilde{n}_{\vb{k}}} \right)^2 \int \frac{d \vb*{\phi}_{\vb{k}}}{ \pi^2} 
e^{-\left(\frac{1 + 2 \tilde{n}_{\vb{k}}}{ \tilde{n}_{\vb{k}}(1 + \tilde{n}_{\vb{k}})}\right)\abs{\phi_{\vb{k}}}^2} e^{-\left(\frac{1 + 2 \tilde{n}_{\vb{k}}}{ \tilde{n}_{\vb{k}}(1 + \tilde{n}_{\vb{k}})}\right)\abs{\phi_{-\vb{k}}}^2} \\ &= \overline{\prod_{\vb{k}}} \frac{1}{1 + 2 \tilde{n}_{\vb{k}}}.
\end{split} 
\label{eq:rhoG2deriv2}
\end{align}
Note that this result could have also been obtained directly from Eq.~\eqref{eq:rhoG}. The important point here is that, in evaluating $\mathrm{Tr} [(\hat{\rho}')^2]$, we encounter integrals with very similar structure as Eq.~\eqref{eq:rhoG2deriv2}, the only difference being the presence of polynomial functions of the coherent state coefficients inside the integrals. Specifically, after following the procedure above we will always be left with a Gaussian integral $\mathcal{I}_{nm}(a)$, that has the form,
\begin{align}
\begin{split}
\mathcal{I}_{nm}(a) &= \frac{1}{\pi} \int_{-\infty}^{\infty} d\mathrm{Re}(\eta) \int_{-\infty}^{\infty} d\mathrm{Im}(\eta) \   \eta^n (\eta^*)^m e^{-a \abs{\eta}^2} \\  &= \frac{1}{4 \pi a \sqrt{a^{n+m}} } \sum_{k = 0}^n \sum_{l = 0}^m i^{k - l}  
\begin{pmatrix} 
    n \\
    k
\end{pmatrix}
\begin{pmatrix} 
    m \\
    l
\end{pmatrix} 
\left(1 + (-1)^{n+m-k-l} \right) \left(1 + (-1)^{k+l} \right) \\ & \qquad \qquad \qquad \qquad \qquad \qquad  \times \Gamma\left(\frac{1 + n+m-k-l}{2} \right) \Gamma\left( \frac{1 + k+l}{2} \right)
\end{split}
\label{eq:GaussInt}
\end{align}
where $a>0$, and $\Gamma(x)$ is the Gamma function. Since it holds that $\mathcal{I}_{nm}(a) = 0$ when $n \ne m$, the only nonzero contributions to $\mathrm{Tr} [(\hat{\rho}')^2]$ are those terms in the momentum sums for which the polynomial function of coherent state coefficients inside the integral is even. For example, when evaluating the trace $\mathrm{Tr} [\hat{\sigma}^2] $, we see from Eq.~\eqref{eq:rhopPhase} that squaring the first term gives a double sum like $\overline{\sum}_{\vb{k} ,\vb{q}} \overline{\sum}_{\vb{k}' ,\vb{q}'}  R_{\vb{k}, \vb{q}}^* R_{\vb{k}', \vb{q}'} \bar{\delta}_{\vb{k}, \vb{q}} \bar{\delta}_{\vb{k}', \vb{q}'} \left[ \hdots \right]$, where $\left[ \hdots \right]$ contains all the integrals over coherent states. Inspection of the polynomial functions inside these integrals with the above condition in mind reveals that the only nonzero contributions to the double sum are those terms for which, 
\begin{align}
\begin{split}
\vb{k}' &= -\vb{k}, \quad \vb{q}' = -\vb{q}, \\
\vb{k}' &= -\vb{k}, \quad \vb{q}' = \vb{k} + \vb{q}, \\ 
\vb{k}' &= -\vb{q}, \quad  \vb{q}' = \vb{k}+\vb{q},
\end{split}
\end{align}
and the swaps $\vb{k}' \leftrightarrow \vb{q}'$. These 6 terms are in fact equivalent due to the inherent permutation symmetries of the $R_{\vb{k}, \vb{q}}$ cumulant. Similarly, squaring the second term in Eq.~\eqref{eq:rhopPhase} gives a double sum like $\overline{\sum}_{\vb{k}} \overline{\sum}_{\vb{k}'}  R_{\vb{k}, \vb{k}}^* R_{\vb{k}', \vb{k}'} \left[ \hdots \right]$, which is nonzero for the term $\vb{k}' = -\vb{k}$. The cross terms always vanish. Finally, taking all the relevant traces using the same method as in Eqs.~\eqref{eq:rhoG2deriv1} and \eqref{eq:rhoG2deriv2} we obtain,
\begin{align}
\begin{split}
\mathrm{Tr} [\hat{\sigma}^2] &= -\frac{4}{3} \left( \overline{\prod_{\vb{p}}}  \frac{1}{1 + 2 \tilde{n}_{\vb{p}}} \right) \overline{\sum_{\vb{k} ,\vb{q}}}  \left(R_{\vb{k},\vb{q}}^*\right)^2  \bar{\delta}_{\vb{k},\vb{q}}    \prod_{\vb{l} = \vb{k}, \vb{q}, -\vb{k} - \vb{q}} \frac{u_{\vb{l}} v_{\vb{l}}^* }{1 + 2\tilde{n}_{\vb{l}}} \\ & \qquad - 4 \overline{\sum_{\vb{k}}}  \left(R_{\vb{k}, \vb{k}}^*\right)^2 \left( \overline{\prod_{\vb{p}}} \frac{1}{1 + 2 \tilde{n}_{\vb{p}}} \right)  \left(\frac{1}{1 + 2 \tilde{n}_{\vb{k}}} \right)^2  u_{\vb{k}}^2 (v_{\vb{k}}^*)^2  \left(\frac{1}{1 + 2 \tilde{n}_{2\vb{k}}} \right) u_{2\vb{k}}v_{2\vb{k}}^*.
\end{split}
\end{align}
Again, the missing terms in the first sum due to $\bar{\delta}_{\vb{k}, \vb{q}}$ are collected in the second sum, such that we can also write,
\begin{align}
\begin{split}
\mathrm{Tr} [\hat{\sigma}^2] &= -\frac{4}{3} \left( \overline{\prod_{\vb{p}}}  \frac{1}{1 + 2 \tilde{n}_{\vb{p}}} \right) \overline{\sum_{\vb{k} ,\vb{q}}}  \left(R_{\vb{k},\vb{q}}^*\right)^2    \prod_{\vb{l} = \vb{k}, \vb{q}, -\vb{k} - \vb{q}} \frac{u_{\vb{l}} v_{\vb{l}}^* }{1 + 2\tilde{n}_{\vb{l}}}.
\end{split}
\end{align}
Following the same procedure, we also derive,
\begin{align}
\begin{split}
\mathrm{Tr} [\hat{\sigma} \hat{\sigma}^{\dagger}] = \frac{1}{6} \left( \overline{\prod_{\vb{p}}}  \frac{1}{1 + 2 \tilde{n}_{\vb{p}}} \right)  \overline{\sum_{\vb{k} ,\vb{q}}}  \abs{R_{\vb{k},\vb{q}}}^2 \prod_{\vb{l} = \vb{k}, \vb{q}, -\vb{k} - \vb{q}} \frac{\abs{u_{\vb{l}}}^2  + \abs{v_{\vb{l}}}^2 }{1 + 2\tilde{n}_{\vb{l}}}.
\end{split}
\end{align}
Taking these results together, we can finally write the trace,
\begin{align}
\begin{split}
\mathrm{Tr} [(\hat{\rho}')^2] &= \frac{1}{3} \left( \overline{\prod_{\vb{p}}}  \frac{1}{1 + 2 \tilde{n}_{\vb{p}}} \right) \mathcal{R},
\end{split}
\end{align}
with,
\begin{align}
\begin{split}
\mathcal{R} = \overline{\sum_{\vb{k} ,\vb{q}}}  \bigg[  \abs{R_{\vb{k},\vb{q}}}^2 \mathcal{U}_{\vb{k}, \vb{q}}    -  \left(R_{\vb{k},\vb{q}}^*\right)^2     \mathcal{V}_{\vb{k}, \vb{q}}  -  \left(R_{\vb{k},\vb{q}}\right)^2     \mathcal{V}_{\vb{k}, \vb{q}}^* \bigg],
\end{split}
\end{align}
where $\mathcal{U}_{\vb{k}, \vb{q}}$ and $\mathcal{V}_{\vb{k}, \vb{q}}$ are defined in Eq.~(9) in the main text. We now return to the non-Gaussianity measure, which becomes,
\begin{align}
\begin{split}
\mathcal{N}(\hat{\rho} || \hat{\rho}_G) &= \frac{1}{6} \frac{ \mathcal{R} }{1 + \frac{1}{3}  \mathcal{R}} \approx \frac{1}{6}  \mathcal{R},
\end{split} 
\label{eq:NG}
\end{align}
where the last equality holds to first order in $\mathcal{R}$, and matches Eq.~(8) in the main text. Finally, it is good to note that the result in Eq.~\eqref{eq:NG} respects the positivity of the non-Gaussianity measure, since it can be rewritten as,
\begin{align}
\begin{split}
\mathcal{N}(\hat{\rho} || \hat{\rho}_G) &= \frac{1}{6}  \overline{\sum_{\vb{k} ,\vb{q}}} \frac{1}{(1 + 2 \tilde{n}_{\vb{k}})(1 + 2 \tilde{n}_{\vb{q}})(1 + 2 \tilde{n}_{-\vb{k} - \vb{q}})} \\ & \qquad  \times \bigg[ \abs{R_{\vb{k}, \vb{q}}^* u_{\vb{k}} u_{\vb{q}} u_{-\vb{k} - \vb{q}} - R_{\vb{k}, \vb{q}} v_{\vb{k}} v_{\vb{q}} v_{-\vb{k} - \vb{q}}}^2 \\
& \qquad  + \abs{R_{\vb{k}, \vb{q}}^* u_{\vb{k}} u_{\vb{q}} v_{-\vb{k} - \vb{q}}^* - R_{\vb{k}, \vb{q}} v_{\vb{k}} v_{\vb{q}} u_{-\vb{k} - \vb{q}}^*}^2  \\
&  \qquad  + \abs{R_{\vb{k}, \vb{q}}^* u_{\vb{k}} v_{\vb{q}}^* u_{-\vb{k} - \vb{q}} - R_{\vb{k}, \vb{q}} v_{\vb{k}} u_{\vb{q}}^* v_{-\vb{k} - \vb{q}}}^2 \\
&  \qquad  + \abs{R_{\vb{k}, \vb{q}}^* v_{\vb{k}}^* u_{\vb{q}} u_{-\vb{k} - \vb{q}} - R_{\vb{k}, \vb{q}} u_{\vb{k}}^* v_{\vb{q}} v_{-\vb{k} - \vb{q}}}^2 \bigg],
\end{split} 
\end{align}
which is always positive. We can calculate this quantity directly from our simulated set of cumulants. In numerical practice, we calculate the normalized non-Gaussianity $\mathcal{N}(\hat{\rho} || \hat{\rho}_G) / V^2$, and then multiply with the square of the actual trap volume in a given experiment for obtaining the characteristic shell width $\Delta k_s^{\mathrm{max}}$, which is plotted in Fig. 3(c) in the main text.

\newpage 
\bibliography{References}